\documentclass[aps,prb,reprint,amsmath,amssymb,superscriptaddress]{revtex4-2}

\usepackage{graphicx}
\usepackage{dcolumn}
\usepackage{bm}
\usepackage{color}

\begin{document}


\title{Magnetic and structural properties of the intermetallic Ce$_{(1-x)}$La$_x$CrGe$_3$ series of compounds}



\author{B. Bosch-Santos}
\email[]{brianna@alumni.usp.br}
\affiliation{NIST Center for Neutron Research, National Institute of Standards and Technology, Gaithersburg, MD 20899, USA}
\affiliation{Instituto de Pesquisas Energ\'{e}ticas e Nucleares – IPEN, S\~{a}o Paulo, S\~{a}o Paulo, SP, 05508-000, Brazil}

\author{G. A. Cabrera-Pasca}
\affiliation{Faculdade de Ci\^{e}ncias Exatas e Tecnologia, Universidade Federal do Par\'{a}, Abaetetuba, PA, 68440-000, Brazil}

\author{E. L. Correa}
\affiliation{Material Measurement Laboratory, National Institute of Standards and Technology - NIST, Gaithersburg, MD 20899, USA}
\affiliation{Instituto de Pesquisas Energ\'{e}ticas e Nucleares – IPEN, S\~{a}o Paulo, S\~{a}o Paulo, SP, 05508-000, Brazil}

\author{B. S. Correa}
\affiliation{Instituto de Pesquisas Energ\'{e}ticas e Nucleares – IPEN, S\~{a}o Paulo, S\~{a}o Paulo, SP, 05508-000, Brazil}

\author{T. N. S. Sales}
\affiliation{Instituto de Pesquisas Energ\'{e}ticas e Nucleares – IPEN, S\~{a}o Paulo, S\~{a}o Paulo, SP, 05508-000, Brazil}

\author{K-W. Moon} 
\affiliation{Material Measurement Laboratory, National Institute of Standards and Technology - NIST, Gaithersburg, MD 20899, USA}

\author{C. L. Dennis}
\affiliation{Material Measurement Laboratory, National Institute of Standards and Technology - NIST, Gaithersburg, MD 20899, USA}

\author{Q. Huang} 
\affiliation{NIST Center for Neutron Research, National Institute of Standards and Technology, Gaithersburg, MD 20899, USA}

\author{J. B. Leao}
\affiliation{NIST Center for Neutron Research, National Institute of Standards and Technology, Gaithersburg, MD 20899, USA}

\author{J. W. Lynn}
\affiliation{NIST Center for Neutron Research, National Institute of Standards and Technology, Gaithersburg, MD 20899, USA}

\author{A. W. Carbonari}
\affiliation{Instituto de Pesquisas Energ\'{e}ticas e Nucleares – IPEN, S\~{a}o Paulo, S\~{a}o Paulo, SP, 05508-000, Brazil}


\date{\today}

\begin{abstract}
The Ce$_{(1-x)}$La$_x$CrGe$_3$ (x = 0, 0.19, 0.43, 0.58 and 1) intermetallic compound system has been investigated by magnetization measurements and neutron scattering techniques to determine the effect of La-doping on the magnetic ordering and exchange interaction between Cr ions.
The structural and magnetic characterization in this series was first verified by X-ray diffraction and bulk magnetization measurements. The samples exhibit the known hexagonal perovskite structure (\textit{P}6$_3$/\textit{mmc} space group) and have a single magnetic phase according to magnetization measurements. In this work, the ferromagnetic ordering temperature for Cr evolves smoothly from a range of 68 K to 77 K for CeCrGe$_3$ to a range of 91 K to 96 K for LaCrGe$_3$ as La replaces Ce. Magnetization results indicate the formation of domain walls below the transition temperature for all the Ce$_{(1-x)}$La$_x$CrGe$_3$ systems investigated. Neutron results indicate  ordered magnetic Cr moments aligned along the \textit{c} axis for the x = 1 LaCrGe$_3$ system, as well as for x = 0.19, 0.43, and 0.58, which contrasts with the x = 0 CeCrGe$_3$ where the moments order in the \textit{ab} plane.

\end{abstract}


\maketitle

\section{\label{Introd}Introduction}
Compounds with rare-earth (RE) and transition metal (TM) elements are interesting magnetic systems due to the possibility of magnetic coupling between the RE and TM atoms. Around fifteen years ago, new intermetallic systems containing RE and TM (RETMGe$_3$ with TM = Ti, Cr) were discovered \cite{Manfrinetti2005, Bie2007} that had an unusual feature: they crystalize in the perovskite-type structure (space group \textit{P}6$_3$/\textit{mmc}) and can exhibit a wide variety of interesting properties including dense Kondo lattice behavior, long range magnetic order, quantum criticality, heavy fermion physics, and superconductivity.

For compounds of this family where the magnetic behavior originates solely from the rare-earth ions, they typically order at low temperatures where Kondo lattice, heavy fermion behavior, and/or superconductivity can emerge, and they are called heavy-fermions compounds. For example, the magnetic ordering of the Ce moment occurs at 5.5 K in CeNiGe$_3$, 14.5 K in CeRhGe$_3$, and 14 K in CeTiGe$_3$.  All of these compounds are heavy fermions \cite{Wang2019, Khan2016}. Furthermore, the TM ions in these compounds have no magnetic moment so that the magnetic behavior originates from the RE atoms \cite{Das2015}. By contrast, in RECrGe$_3$ the Cr atoms exhibit magnetic moments which order ferromagnetically at much higher temperatures, as indicated by the end member LaCrGe$_3$ of particular interest here where RE magnetic moments are absent \cite{Bie2007}.  One of the interesting aspects of this system is that the ferromagnetism can be suppressed toward a purported quantum critical point (QCP) by doping, \cite{Lin2013} pressure, or applied magnetic field  \cite{Taufour2018, Taufour2016}, but the system successfully evades this scenario in an interesting way. With the introduction of magnetic rare earths the emergence of magnetic interactions between the Cr atoms and the 4\textit{f}-RE moments presents an interesting situation for this strongly correlated electron system. In these compounds the ground state is ruled to a large extent by two competitive interactions: Ruderman-Kittel-Kasuya-Yosida (RKKY) indirect exchange, and Kondo exchange \cite{Das2016,Bie2007}.

Specifically, LaCrGe$_3$ and CeCrGe$_3$ are two interesting end members of this family due to the fact that Ce has one 4\textit{f} electron and La has no 4\textit{f} electron. Furthermore, previous reports have shown that LaCrGe$_3$ has a fragile magnetism and CeCrGe$_3$ is a moderate heavy-fermion system \cite{Das2014, Nguyen2018}. A neutron diffraction investigation of the fragile ferromagnet LaCrGe$_3$ shows that the Cr moments are aligned along the \textit{c} axis below the ordering temperature of 78 K, but have a spin-canted structure below 3 K. At 1.7 K, the spins form an angle of 32$^{\circ}$ with respect to the c axis \cite{Cadogan2013}. On the other hand, the alignment of Cr moments in the moderate heavy-fermion CeCrGe$_3$ is still under discussion; previous work has shown the possibility of analyzing the same neutron diffraction data using two very different assumptions, Cr spins parallel to the \textit{ab} plane, or parallel to the \textit{c} axis direction. Results from both models are quite different and, therefore, further investigation is necessary. The presence of Ce magnetic moments in CeCrGe$_3$ can induce both the Ce and the Cr moments to be aligned in the \textit{ab} plane in the ferromagnetic phase, as shown by neutron diffraction \cite{Das2016}. Controversially, a study using muon spin relaxation has not shown ordering of the Ce moments \cite{Das2015}. Finally, an interesting characteristic of CeCrGe$_3$ is that it has been shown to present the Kondo effect with heavy-fermion  behavior, where Ce atoms have an integral +3 valence state \cite{Das2014}.

Previous work has presented several values for the magnetic transition temperature. The ferromagnetic transition temperature in LaCrGe$_3$ is reported to vary from 78 K to 88 K \cite{Bie2007,Cadogan2013,Das2014,Taufour2016}, whereas for CeCrGe$_3$, this temperature can vary from 66 K to 73 K \cite{Bie2007,Das2014}. Furthermore, there are claims of the existence of a tricritical point near 40 K for LaCrGe$_3$ when external pressure is applied\cite{Taufour2016}.

Despite the interest in this family of compounds, the understanding of these materials remains elusive. CeCrGe$_3$ and LaCrGe$_3$ have been reported to present different collinear alignments of the Cr spins: in LaCrGe$_3$ the alignment is along the \textit{c} axis whereas in CeCrGe$_3$ it is an unresolved issue \cite{Cadogan2013,Das2016}. These issues, including the Cr spin alignment in CeCrGe$_3$, the possible difference in its alignment in CeCrGe$_3$ and LaCrGe$_3$, and the possible ordering of the Ce moments, have motivated us to investigate the La doping in the CeCrGe$_3$ to ascertain the influence of the 4\textit{f} electron in the magnetic behavior of doped compounds. In addition, the fragile magnetism of LaCrGe$_3$ appears to be due to the Cr \textit{d} orbital possessing a peak just below the Fermi level. Modest changes in pressure can then cause substantial changes in the magnetic characteristics \cite{Nguyen2018}. This pressure can be a chemical pressure or physical pressure; in this work the La substitution by Ce will cause a decrease in the crystal cell volume.
Another interesting aspect that was not well-studied, and reported for the first time by Bie, \textit{et al.}, is the magnetic domain formation in the family of compounds RECrGe$_3$ (RE = La - Nd, Sm) \cite{Bie2007}. Afterwards Lemoine, \textit{et al.} investigated the magnetic domain formation and domain wall movement in NdCrGe$_3$ \cite{Lemoine}. These are important properties to be investigated due to there influence on the magnetic transition temperature, magnetic moment and spin alignment \cite{Lemoine, Binder, Bocarsly, Nehla}.

In this work, we have combined neutron scattering techniques and magnetization measurements to investigate the magnetic properties of the series Ce$_{(1-x)}$La$_x$CrGe$_3$ (x = 0, 0.19, 0.43, 0.58 and 1). Results from the magnetization measurements revealed the formation of domain walls for all compounds in this series. Moreover, neutron powder diffraction demonstrated that the Cr moments align along the \textit{c} axis for the x = 0.19, 0.43, 0.58 and 1 and align in the \textit{ab} plane for x = 0. Here we discuss the contrast in the alignment direction of the Cr moments, the large difference in the values of effective magnetic moment and ordered magnetic moment, as well as the divergence between the temperature transition values determined using magnetization measurements and neutron diffraction.

\section{\label{Exp}Experimental Details}
The intermetallic Ce$_{(1-x)}$La$_x$CrGe$_3$ (x = 0, 0.19, 0.43, 0.58, 1) compounds were prepared by arc melting in an argon atmosphere. Starting elements of La and Ce pieces with 99.9 \% of purity and Cr and Ge pieces with 99.999\% of purity, were added in the stoichiometric ratio. After melting, the resulting ingot of each sample was sealed in an evacuated quartz tube ($\sim 10^{-2}Pa$), which was then annealed at 900 $^{\circ}$C for 2 weeks. After annealing, the structural quality of the samples was verified by x-ray diffraction (XRD) [Rigaku Ultima III] with Cu-K$\alpha$ radiation. The CeCrGe$_3$, LaCrGe$_3$ and Ce$_{0.58}$La$_{0.43}$CrGe$_3$ samples were single phase corresponding to the expected hexagonal perovskite-type structure with space group \textit{P}6$_3$/\textit{mmc} as previously reported \cite{Bie2007}. Ce$_{0.42}$La$_{0.58}$CrGe$_3$ and Ce$_{0.81}$La$_{0.19}$CrGe$_3$ samples showed an additional small impurity phase ($<$ 1 \% and around 6 \%, respectively). The impurity phase was identified as La(Ce)Ge$_2$ crystallized in the tetragonal structure with space group \textit{I}4$_1$/\textit{amd}. This phase forms at 1500 $^{\circ}$C and is a common impurity found in these types of compounds when prepared by arc melting \cite{Eremenko1971}. All intermetallic samples were prepared in the Hyperfine Interactions Laboratory at the Nuclear and Energy Research Institute (IPEN), except the Ce$_{0.8}$La$_{0.2}$CrGe$_3$ sample that was prepared at National Institute of Standard and Technology (NIST).

In order to investigate the macroscopic magnetic properties, samples were magnetically characterized using a Superconducting Quantum Interference Device Vibrating-Sample Magnetometer (SQUID-VSM, Quantum Design). The characterization was performed after cooling the sample in zero field, by measuring the DC magnetic moment (M) at $\mu_0$H = 0.01 T while warming followed by cooling. The first curve, after the zero field cool, is referred to as the ZFC data.  The second curve, measured while cooling under an applied field, is referred to as the FC data. The M vs T was then converted to ${\chi_{DC}(T)}$ where ${\chi_{DC} = M/H}$. In addition, the AC susceptibility (both real ($\chi$') and imaginary ($\chi$'') components) as a function of temperature  for different frequencies \textit{f} = 1 Hz, 10 Hz and 100 Hz and AC field of 10$^{-4}$ T, was measured. These were performed to cross-check the features observed in the DC magnetization measurements and to discard any other possible magnetic phase occurring in the system due to the La-doping.

To study the magnetic structure and better understand the magnetic behavior in this series, neutron diffraction data were obtained at the NIST Center for Neutron Research (NCNR). For both neutron powder diffraction (NPD) measurements using high resolution or coarse resolution/high intensity, the mass of the polycrystalline powder samples used in the measurements was around 1.3 g for Ce$_{(1-x)}$La$_x$CrGe$_3$ (x = 0, 0.43, 0.58, 1) and $\sim$ 4 g for the Ce$_{0.81}$La$_{0.19}$CrGe$_3$. Each sample was sealed in a vanadium container under helium atmosphere in a glove box. 
NPD data were collected using the BT-1 32-detector High-Resolution Neutron Powder Diffractometer at NCNR \cite{Santoro2001} over the range 2$\theta$ = 3$^{\circ}$ to 2$\theta$ = 168$^{\circ}$ with a step size of 0.050$^{\circ}$. For all samples, measurements were taken at 5 K, 295 K and at a temperature slightly above the magnetic transition temperature determined by magnetization measurements. Temperature dependent data sets were obtained using a closed cycle refrigeration system. A Cu (311) monochromator (wavelength $\lambda$ = 1.5402 (2) {\AA}) with a take-off angle of 90$^{\circ}$ and 60 minutes of arc in-pile collimation was used to collect room temperature data. A Ge (311) monochromator ($\lambda$ = 2.0787(2) {\AA}) and in-pile collimation of 60 minutes of arc was used for the low temperature measurements due to its higher resolution at low angles. Rietveld refinement was used to solve the nuclear and magnetic structure for all patterns and it was performed using Generalized Structure Analysis System (GSAS) software \cite{GSAS}.
The temperature dependence of the magnetic Bragg peaks and the order parameters were obtained from measurements at the BT-7 triple-axis spectrometer \cite{Lynn2012} at NCNR. Measurements were carried out using a pyrolytic graphite [PG (002)] monochromator with wavelength 2.359 {\AA}. The data were taken using  80' to 80' open collimation before the sample, and an 80' radial collimator in front of the position sensitive detector, and scattering angles of 2$\theta$ = 13$^{\circ}$ to 2$\theta$ = 103$^{\circ}$. Samples were measured in powder form inside a vanadium can under low pressure He gas. Measurements were performed in a closed cycle refrigerator (CCR) with a range of temperatures from 2 K to 120 K for Ce$_{(1-x)}$La$_x$CrGe$_3$ (x = 0.19, 0.43, 0.58, 1). CeCrGe$_3$ was first measured in a temperature range of 2 K to 10 K to verify the magnetic transition from the Ce spins, and then from 10 K to 120 K. The magnetic peak scattering was located in the (100) plane for all samples. All BT-7 data were analyzed using Data Analysis and Visualization Environment - DAVE software \cite{DAVE}.
The error bars in Fig. \ref{fig9} (Section IV - Discussion) are the standard deviations of the adjusted parameters from GSAS software. For lattice parameters the error is 0.001 Å and for unit cell volume the error is 0.1 Å$^3$, as shown in Table \ref{tab2}.

\section{\label{Exp}Experimental Results}

\subsection{\label{Mag}Magnetization measurements}
\begin{figure}[ht!]
\includegraphics [scale = 0.23]{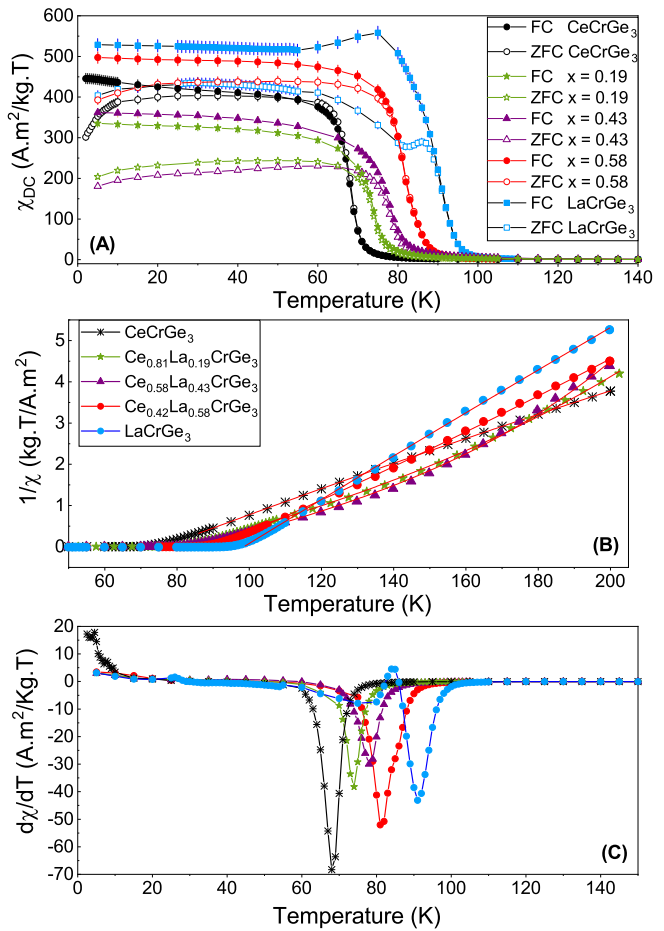}
\caption{(A) FC and ZFC $\chi_{DC}$ for Ce$_{(1-x)}$La$_x$CrGe$_3$ (x = 0, 0.19, 0.43, 0.58, 1). (B) Inverse of $\chi_{DC}$ as a function of temperature. Solid lines are the modified Curie-Weiss law fit. (C) Derivative of DC susceptibility as a function of temperature. All error bars are shown and represent 1 $\sigma$.  However, the error bars may be smaller than the symbol.}
\label{fig1}
\end{figure}

Fig. \ref{fig1} displays the magnetization measurement results. For $\chi_{DC}$(T) all the samples show a clear ferromagnetic-paramagnetic transition. The ZFC and FC data demonstrate a clear turn-up after which $\chi_{DC}$(T) increases rapidly with decreasing temperature (see Fig. \ref{fig1} (A)). Also, in Fig. \ref{fig1} (A), a divergence between ZFC and FC curves below the temperature transition can be seen. This behavior is likely due to magnetic domains which indicate a strong magnetocrystalline anisotropy, as has been presented by Bie \textit{et al.} in RECrGe$_3$ (RE = La - Nr, Sm) \cite{Bie2007} and Lemione \textit{et al.} in NdCrGe$_3$ \cite{Lemoine}. Additionally, LaCrGe$_3$ in the ZFC curve presents a little bump around 80 K suggesting a possible spin reorientation or the coexistence of both antiferromagentic and ferromagnetic phases.

\begin{table*}
\caption{\label{tab1}%
Results obtained by fitting the modified Curie-Weiss law to $\chi^{-1}$ for CeCrGe$_3$, Ce$_{0.81}$La$_{0.19}$CrGe$_3$ (x = 0.19), Ce$_{0.58}$La$_{0.43}$CrGe$_3$ (x = 0.43), Ce$_{0.42}$La$_{0.58}$CrGe$_3$ (x = 0.58), and LaCrGe$_3$: Curie Temperature ($T_C^{(DC)}$ and $T_C^{(AC)}$), Curie-Weiss temperature ($\theta_{CW}$), and effective magnetic moment ($\mu_{eff}$). Values of $\mu_{eff}(y)$ are the estimate $\mu_{eff}$ values considering Ce$^{3+}$ and Cr$^{3+}$ ions and the Ce percentage in each one sample. $T_C$ (lit. $T_C$) and $\mu_{eff}$ (lit. $\mu_{eff}$) from literature are also displayed. The numbers between parentheses are shown the error bars and represent 1 $\sigma$.}
\begin{ruledtabular}
\begin{tabular}{lcccccccc}
\textrm{Sample} &
\textrm{$T_C^{(DC)}$} &
\textrm{$T_C^{(AC)}$} &
\textrm{$\theta_{CW}$} &
\textrm{C} &
\textrm{$\mu_{eff}$}&
\textrm{$\mu_{eff}(y)$}&
\textrm{lit. $T_C$} &
\textrm{lit. $\mu_{eff}$}
\\
    &   \textrm{(K)} & \textrm{(K)} & \textrm{(K)} & \textrm{(K.A/T.m)}  &\textrm{($\mu_B$)}  &\textrm{($\mu_B$)} & \textrm{(K)} & \textrm{($\mu_B$)}\\
\hline
CeCrGe$_3$        & 68.0(1) & 68.6(4) &  74.4(2)  &  32.1(3)   & 3.23(2) & 4.57 & 73\footnotemark[1], 70\footnotemark[2] & 3.18\footnotemark[1], 3.36\footnotemark[2]\\
\textrm{x} = 0.19 & 73.8(1) & 73.8(3) & 78.8(3)  & 43.6(6)  & 3.77(3) & 4.43  & --- & ---\\
\textrm{x} = 0.43 & 77.8(1) &  78.2(2) & 84.1(5)  & 48.8(3)    & 4.00(5) & 4.26  & --- & ---\\
\textrm{x} = 0.58 & 81.8(1) & 83.8(2) & 90.5(2)  & 26.7(4)    & 2.94(2) & 4.14  & --- & ---\\
LaCrGe$_3$        & 91.4(1) & 96.8(8) & 98.6(2)   & 18.5(2)  & 2.46(2) & 3.8 & 86\footnotemark[3], 88\footnotemark[1] & 2.38\footnotemark[3], 2.66\footnotemark[1]\\
\end{tabular}
\end{ruledtabular}
\footnotetext[1]{reported by Das \textit{et al.} \cite{Das2014}}
\footnotetext[2]{reported by Das \textit{et al.} \cite{Das2015}}
\footnotetext[3]{reported by Taufour \textit{et al.} \cite{Taufour2018}}
\end{table*}

\begin{figure}[ht!]
\includegraphics [scale=0.32]{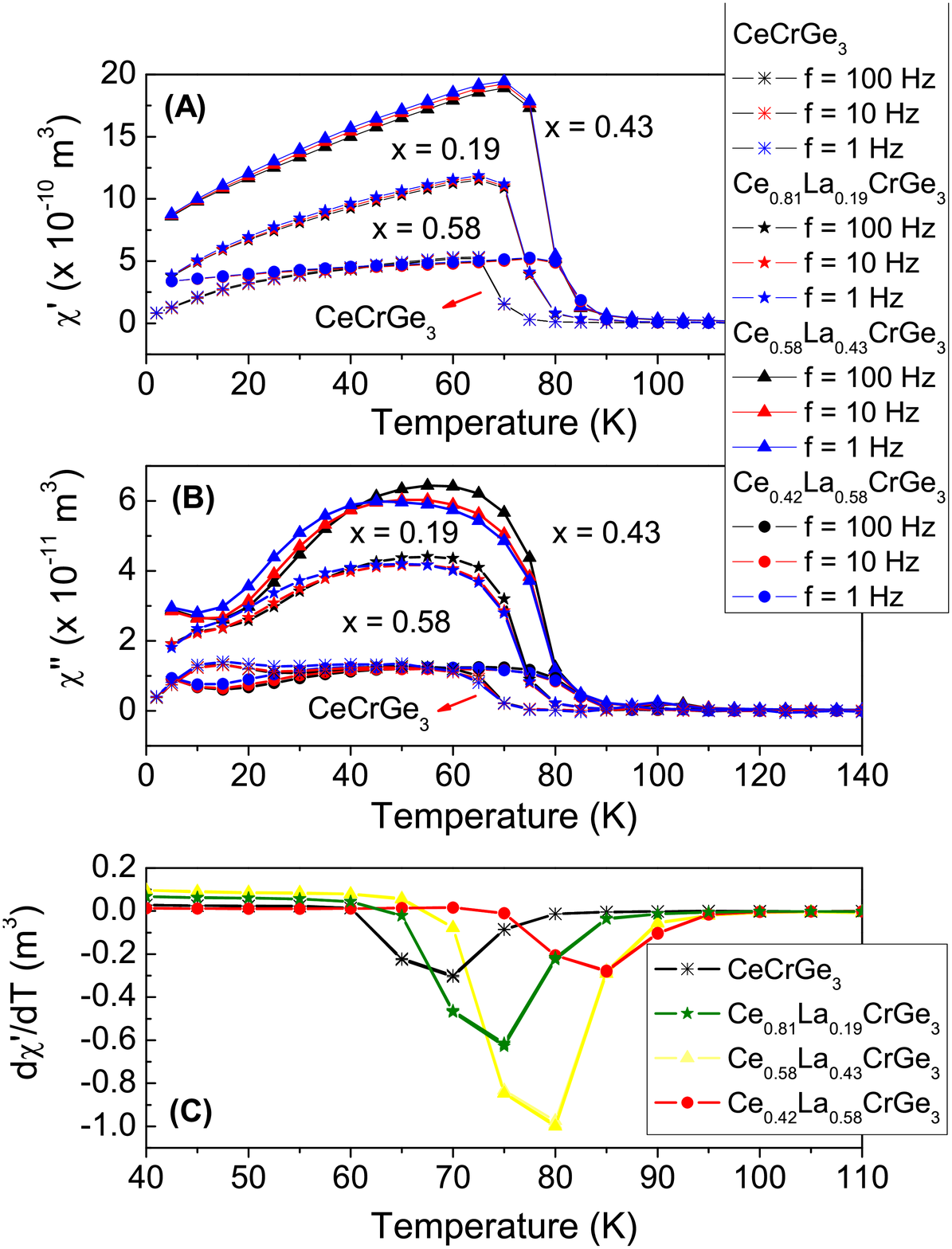}
\caption{$\chi_{AC}$ for Ce$_{(1-x)}$La$_x$CrGe$_3$ (x = 0, 0.19, 0.43, 0.58) (A) real component ($\chi$') (B) imaginary component ($\chi$'') and (C) Derivative of $\chi_{AC}$  real component as a function of temperature. All error bars are shown and represent 1 $\sigma$.  However, the error bars may be smaller than the symbol.}
\label{figACS}
\end{figure}

\begin{figure}[ht!]
\includegraphics [scale=0.32]{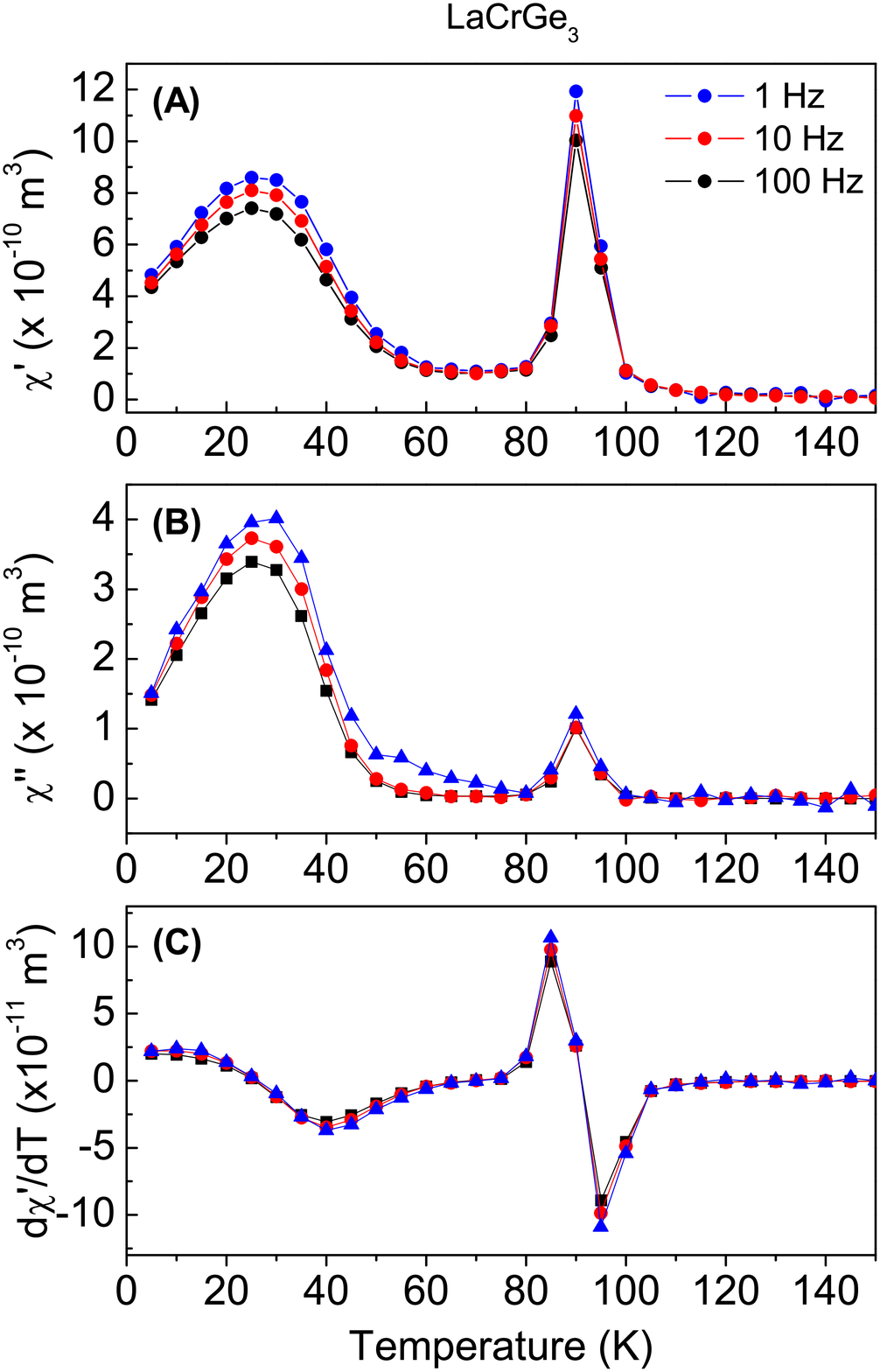}
\caption{$\chi_{AC}$ for LaCrGe$_3$ (A) real component ($\chi$') (B) imaginary component ($\chi$'') and (C) Derivative of $\chi_{AC}$  real component as a function of temperature. All error bars are shown and represent 1 $\sigma$.  However, the error bars may be smaller than the symbol.}
\label{figACS_La}
\end{figure}

Above the magnetic transition temperature (in the paramagnetic phase) $\chi_{DC}$(T) shows the characteristic Curie-Weiss behavior for paramagnetic states. In the case of doped samples (x = 0.19, 0.43 and 0.58) the inverse $\chi_{DC}(T)$ data have shown a non-linear behavior that present a smooth curvature (see Fig.  \ref{fig1} (B)). Hence, the inverse of modified Curie-Weiss law was used to fit the data, described by $\chi = C/(T-\theta_{CW}) + \chi_0$ and fit in the paramagnetic region. Here $C$ is the Curie constant, $\theta_{CW}$ is the Curie-Weiss temperature and $\chi_0$ is the temperature independent susceptibility \cite{Amarotti1984}. From the fit of $\chi^{-1}$ (see Fig. \ref{fig1} (B)), $C$ and $\theta_{CW}$ can be calculated and the results are used to determine the effective magnetic moment given by $\mu_{eff} = (2.828 \times \sqrt{C \times MM})\mu_B$, where $MM$ is the molecular mass of each compound.
The ferromagnetic ordering temperature ($T_C^{(DC)}$) values were determine by the minimum of the  derivative (d$\chi$/dT) of the ZFC curve (see Fig. \ref{fig1} (C)). The values obtained from the fits, as well as the $T_C^{(DC)}$ of different quantities of La-doping, are shown on Table \ref{tab1}. Note that in order to obtain the ferromagnetic ordering temperature ($T_C^{(AC)}$) from $\chi_{AC}$(T) data, we have used the minimum of the derivative (d$\chi_{AC}$/dT) of the real part ($\chi$') curve (see Fig. \ref{figACS} (C) and Fig. \ref{figACS_La} (C)).

As shown in Table \ref{tab1}, $\mu_{eff}$ displays a peak, with the values for the doped samples being higher than the values for pure samples (CeCrGe$_3$ and LaCrGe$_3$). To estimate the value of $\mu_{eff}$ in the doped samples, considering the Ce contribution, we used the equation $\mu_{eff} (y) = \sqrt{(\mu_{Ce^{3+}})^{2}(y) + (\mu_{Cr^{3+}})^{2}} \mu_B$, where y is the Ce quantity. 
We are not considering Cr$^{4+}$ because, from data of density of state (DOS) curves, the compound LaCrGe$_3$ presents the general formulation (La$^{3+}$)(Cr$^{3+}$)(Ge$^{2-}$)$_3$ \cite{Bie2007}. Furthermore, we assume in our samples that Ce is trivalent because Ce$^{4+}$ has the 4$f^0$ orbital and therefore has no magnetic contribution. Thus, from previous X-ray absorption spectroscopy results, we use the following values Cr$^{3+}$ (3.8 $\mu_{B}$) and Ce$^{3+}$ (2.54 $\mu_{B}$) \cite{Das2014}. The estimated values of $\mu_{eff}$ are displayed on Table \ref{tab1}.

In Table \ref{tab1}, the positive values of Curie-Weiss temperatures for all samples confirm net ferromagnetic interactions in this system. Previous works have reported $T_C$  and $\mu_{eff}$  results from magnetization measurements for LaCrGe$_3$ and CeCrGe$_3$ \cite{Das2014, Das2015, Taufour2018}. Values of $T_C^{(DC)}$ and $T_C^{(AC)}$ show a displacement when compared to those reported in the literature as can be seen in Table \ref{tab1}, except $T_C^{(AC)}$ for CeCrGe$_3$ that is in good agreement with that from Das et al. \cite{Das2015}. We do not have enough data to explain this discrepancy. Overall, values of $\mu_{eff}$ agree reasonably well with those from the literature. Interestingly, the calculated values for $\mu_{eff}$ are linear with La doping, but the measured values are not.  This does imply that there is another factor involved which has not yet been accounted for. Since the Curie-Weiss fits are in the paramagnetic phase we can expect contributions from the magnetic moments but these moments will not necessarily couple at low temperature. Therefore, the $\mu_{eff}$ found using the Curie-Weiss law should not be higher than the $\mu_{eff}$ for the free ion, as was the case.

$\chi_{AC}$(T) (both ($\chi$') and ($\chi$'') component) curves are shown in Fig. \ref{figACS} (A and B) for Ce$_{(1-x)}$La$_x$CrGe$_3$ (x = 0, 0.19, 0.43 and 0.58) and Fig. \ref{figACS_La} (A and B) for LaCrGe$_3$. As can be seen, for all samples there is a clear upturn in the $\chi$' component with decreasing temperature, characteristic of the paramagnetic-ferromagnetic transition. Below this transition, the $\chi_{AC}$(T) (($\chi$') and ($\chi$'')) curves for the CeCrGe$_3$ and doped samples (x = 0.19, 0.43 and 0.58) display a typical ferromagnetic behavior, with a smooth decrease of the magnetic susceptibility below T$_C$ (see Fig. \ref{figACS} (A and B)). This decrease in $\chi$' usually occurs due to a reduction in the ability of the material to respond to the low ac magnetic field (here 10$^{-14}$ T) \cite{Levin}. The decrease in $\chi$'' likely occurs due to the formation of domain walls and magnetic domains within the material, that absorb energy when they are pinned while trying to grow/shrink \cite{Levin}.  This is most evident in the x = 0, x = 0.19, and x = 0.43 samples.

On the other hand, for LaCrGe$_3$ the behavior of $\chi$' is related to the magnetic domain structure, which appears right below $T_C$. With the decrease in temperature, it is clear that the material loses its ability to respond to the AC magnetic field. For example, if multiple pinned domains form just below T$_C$ and the excitation field is not large enough to shift the moment, then the AC response can shift to near zero. In $\chi$'' a similar peak can be seen in the same temperature range, corroborating the formation of magnetic domains, which absorb energy while trying to move them. Also, at lower temperature (20-30 K), a second larger peak can be seen in both $\chi$' and $\chi$'' further supporting this explanation. Alternatively, the second peak in $\chi$'' could be related to spin-glass behavior. However, despite the frequency dependence, a change in the peak position as a function of the temperature is not observed \cite{Aslibeiki}.

\subsection{\label{Neutron}Neutron measurements}
\subsubsection{Neutron Powder Diffraction at BT-1}

\begin{figure}[ht!]
\includegraphics [scale=0.35]{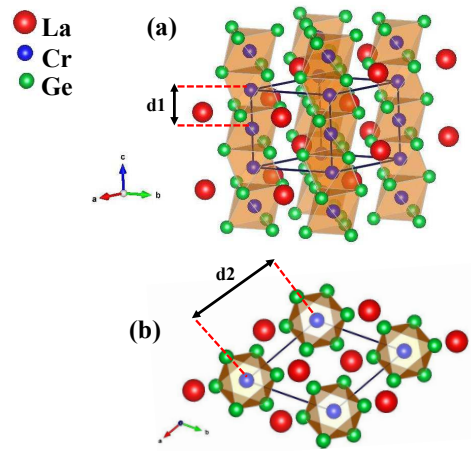}
\caption{Crystallographic structure of hexagonal perovskite-type (\textit{P}6$_3$/\textit{mmc} space group) RCrGe$_3$ (R = La, Ce) viewed (A) perpendicular to the c-direction in a polyhedral representation (B) in the ab plane. d1 and d2 are the intralayer and interlayer distances, respectively.}
\label{fig2}
\end{figure}

\begin{figure}[ht!]
\includegraphics [scale=0.33]{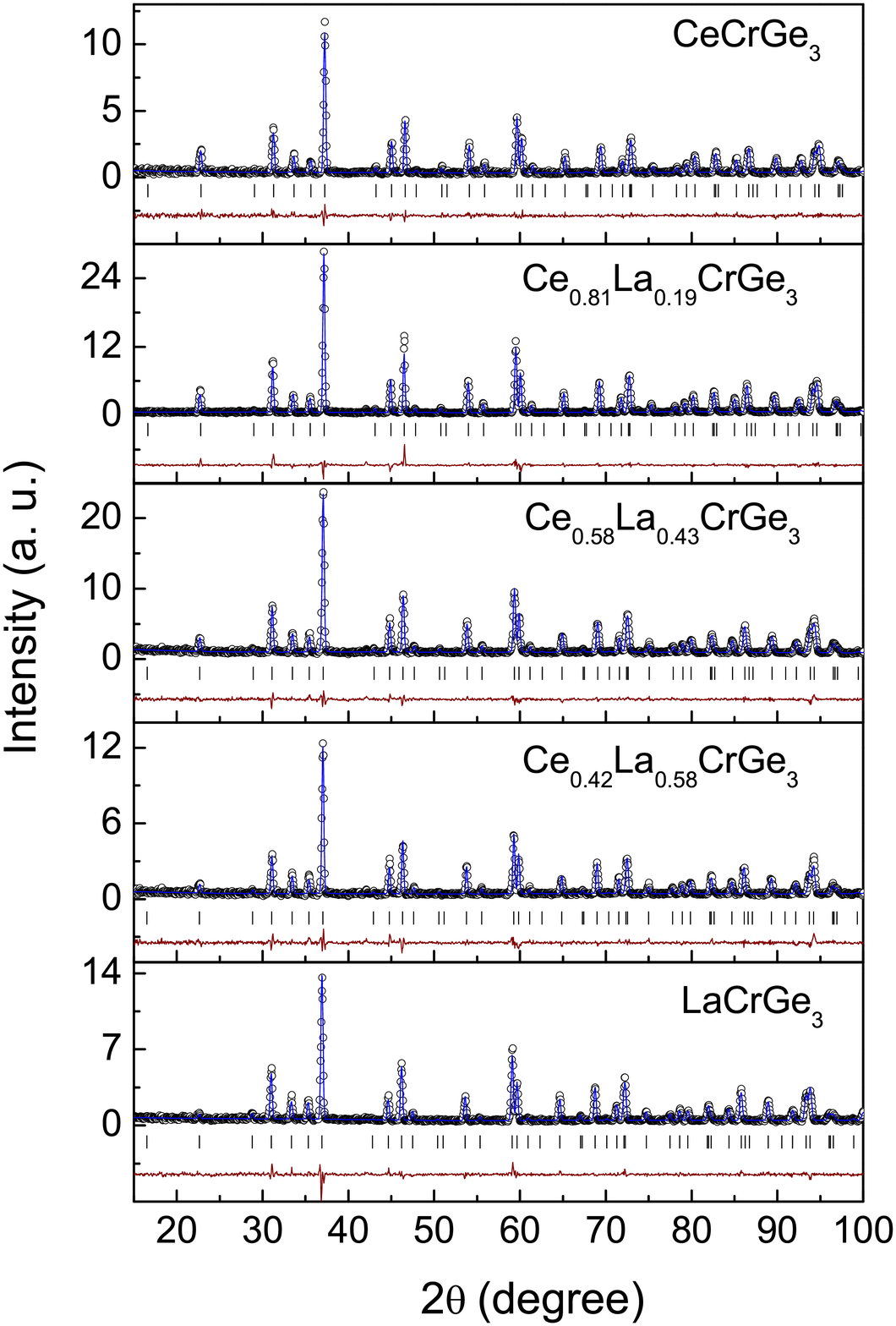}
\caption{NPD results with Rietveld refinement for Ce$_{(1-x)}$La$_x$CrGe$_3$ (x = 0, 0.19, 0.43, 0.58, 1) at room temperature. Blue continuous lines are the Rietveld refinement fit, red lines represent the difference between the observed and calculated intensities and the small vertical black lines indicate the theoretical peaks position of the \textit{P}6$_3$/\textit{mmc} space group. All error bars are shown and represent 1 $\sigma$.  However, the error bars may be smaller than the symbol.}
\label{fig3}
\end{figure}

\begin{table*}
\caption{\label{tab2}%
Structural parameters obtained from the neutron Rietveld refinement analysis of measurements taken at 295 K on Ce$_{(1-x)}$La$_{x}$CrGe$_3$ (\textit{x} = 0, 0.19, 0.43, 0.58, 1) series: unit cell parameters (\textit{a} and \textit{c}), Ge atom position ($x_{Ge}$ and $y_{Ge}$) and chi-squared (quality of the fit)  ($\chi^2$).}
\begin{ruledtabular}
\begin{tabular}{lccccc}
         & \textrm{CeCrGe$_3$} 
         & \textrm{Ce$_{0.81}$La$_{0.19}$CrGe$_3$}
         & \textrm{Ce$_{0.58}$La$_{0.43}$CrGe$_3$}
         & \textrm{Ce$_{0.42}$La$_{0.58}$CrGe$_3$}
         & \textrm{LaCrGe$_3$}
\\
\colrule
\textrm{$a$ ({\AA})} & 6.140(1)   & 6.153(1)  & 6.165(1)  & 6.168(1)  &  6.176(1)  \\
\textrm{$c$ ({\AA})} & 5.711(1)   & 5.723(1)  & 5.735(1)  & 5.737(1)  &  5.745(1) \\
\textrm{Occ(Ce) }    &    ---     & 0.81(2)   & 0.58(2)   & 0.42(1)   &  ---       \\
\textrm{Occ(La) }    &    ---     & 0.19(2)   & 0.43(2)   & 0.58(1)   &  ---        \\
\textrm{$x_{Ge}$ }   & 0.1928(1)  & 0.1930(1) & 0.1931(1) & 0.1931(1) & 0.1930(1)\\
\textrm{$y_{Ge}$ }   & 0.3856(1)  & 0.3861(1) & 0.3863(1) & 0.3862(1) & 0.3860(2)\\
\textrm{$\chi^2$}    & 0.8889     &  1.30     &  0.9609   & 1.092     & 0.930 \\ 
\end{tabular}
\end{ruledtabular}
\end{table*}

Similar to what is observed with XRD measurements, CeCrGe$_3$, LaCrGe$_3$ and Ce$_{0.58}$La$_{0.43}$CrGe$_3$ exhibit a single phase of the expected hexagonal perovskite type structure with space group \textit{P}6$_3$/\textit{mmc}, without observable impurities. On the other hand, Ce$_{0.42}$La$_{0.58}$CrGe$_3$  and Ce$_{0.81}$La$_{0.19}$CrGe$_3$ show less than 1 \% and around 6 \% of a secondary phase, respectively. In both cases the secondary phases were identified as the La(Ce)Ge$_2$ tetragonal phase with space group \textit{I}4$_1$/\textit{amd}. As explained in related papers \cite{Das2016,Cadogan2013}, the appearance of this impurity in these compounds is common due to the high temperature used in the arc melting. Nevertheless, such secondary phases are non-magnetic and their atomic peak positions do not interfere with any magnetic peak positions, and thus have no influence on the magnetic refinements.

Figure \ref{fig2} shows the crystallographic structure from the NPD pattern recorded in the paramagnetic state (at room temperature). The structure consists of octahedral chains formed by Ge face-sharing Cr-centers separated by RE atoms, and the Cr sublattice has a cubic structure. The RE atoms occupy the 2d site (1/3, 2/3, 3/4), Cr atoms occupy the 2a site (0, 0, 0) and Ge atoms occupy the 6h site (x, 2x, 1/4) \cite{Bie2007}. d1 and d2 are the intralayer (distance between two consecutive Cr atoms in the \textit{c} axis) and interlayer (distance between two consecutive Cr atoms in the \textit{ab} plane) distances, respectively.
Results at room temperature are presented in Table \ref{tab2} and the fit to the diffraction patterns are shown in Fig. \ref{fig3}.

\begin{figure}
\includegraphics [scale=0.35]{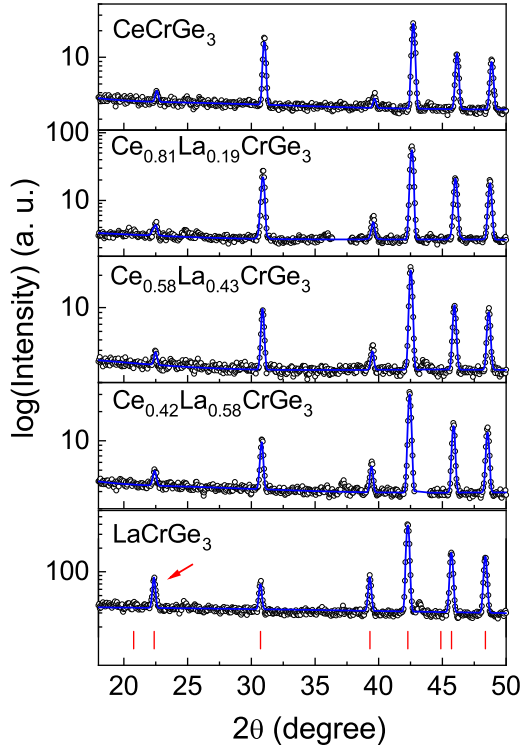}
\caption{NPD patterns of Ce$_1-x$La$_x$CrGe$_3$ (x = 0, 0.19, 0.43, 0.58, 1) at 5 K collected on BT1. Vertical red lines indicate the theoretical magnetic peak positions with hexagonal \textit{P}6$_3$/\textit{mm'c'} symmetry (Cr spin align to the c axis). Arrow for the LaCrGe$_3$ pattern points to the FM peak at 2$\theta$ = 22.5$^{\circ}$. All error bars represent 1 $\sigma$ and are shown, but they may be smaller than the symbol.}
\label{fig4}
\end{figure}

\begin{figure}
\includegraphics [scale=0.3]{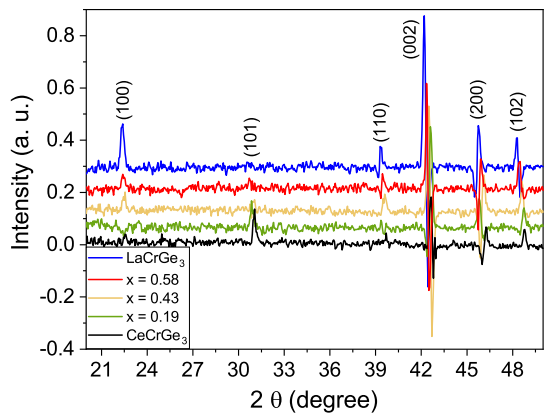}
\caption{Normalized magnetic intensity peaks measured at 5 K for CeCrGe$_3$, Ce$_{0.81}$La$_{0.19}$CrGe$_3$, Ce$_{0.58}$La$_{0.43}$CrGe$_3$, Ce$_{0.42}$La$_{0.58}$CrGe$_3$ and LaCrGe$_3$. All error bars represent 1 $\sigma$ and are shown, but they may be smaller than the symbol.}
\label{fig5}
\end{figure}

NPD results for LaCrGe$_3$ measured at 5 K reveal a significant intensity enhancement for the (100) Bragg reflection when compared to that in the paramagnetic state at 110 K. This ferromagnetic peak at {2$\theta$ = 22.5$^{\circ}$} is the strongest magnetic peak (see Fig. \ref{fig4}). In addition, we have a small intensity enhancement for the (110) and (102) Bragg reflections. Finally, it is clear that there is a small shift in the positions of the (002) and (200) reflections due to the thermal variation of the lattice parameter and Debye-Waller factor and possibly a small magnetic contribution (see Fig. \ref{fig5}). All the magnetic intensities were found in the same position as the nuclear peaks,  as expected for a ferromagnetic structure. The Cr spins are aligned along the \textit{c} axis direction forming a pseudo-1D structure. Structural parameters for this sample are shown in Table \ref{tab3} and present an ordered magnetic moment of 1.40(5) $\mu_B$, which is due to the magnetic Cr atoms. This value is in good agreement with 1.31(4) $\mu_B$ at 3 K for Cr spins aligned along the \textit{c} axis from reference \cite{Cadogan2013}. Lattice parameters for LaCrGe$_3$ (\textit{a} = 6.172(1) {\AA} and \textit{c} = 5.742(1) {\AA} at 120 K and \textit{a} = 6.166(1) {\AA} and \textit{c} = 5.749(1) {\AA} at 3K) are in good agreement with those presented in previous studies \cite{Cadogan2013}.

\begin{table*}
\caption{\label{tab3}%
Structural and magnetic parameters obtained from the Rietveld refinement analysis of measurements taken at 5 K for the Ce$_{(1-x)}$La$_{x}$CrGe$_3$ (\textit{x} = 0, 0.19, 0.43, 0.58, 1) series: unit cell parameters (\textit{a} and \textit{c}), Ge atom position ($x_{Ge}$ and $y_{Ge}$) and chi-squared (quality of the fit)  ($\chi^2$). Cr ion magnetic moment ($\mu_{Cr}$) along the \textit{c} axis for x = 0.19, 0.43, 0.58 and 1, and for CeCrGe$_3$ in both directions along the \textit{c} and parallel to the \textit{ab} plane.}.
\begin{ruledtabular}
\begin{tabular}{lccccc}
         & \textrm{CeCrGe$_3$} 
         & \textrm{Ce$_{0.81}$La$_{0.19}$CrGe$_3$}
         & \textrm{Ce$_{0.58}$La$_{0.43}$CrGe$_3$}
         & \textrm{Ce$_{0.42}$La$_{0.58}$CrGe$_3$}
         & \textrm{LaCrGe$_3$}
\\
\colrule
\textrm{$a$ ({\AA})} & 6.112(1)  & 6.129(1)  & 6.140(1)  & 6.151(1)  &  6.169(1) \\
\textrm{$c$ ({\AA})} & 5.696(1)  & 5.712(1)  & 5.723(1)  & 5.736(1)  &  5.752(1)  \\
\textrm{Occ(Ce) }    &    ---    & 0.82(2)   & 0.58(2)   & 0.42(1)   &  ---       \\
\textrm{Occ(La) }    &    ---    & 0.18(2)   & 0.43(2)   & 0.58(1)   &  ---        \\
\textrm{$x_{Ge}$ }   & 0.1928(1)\footnotemark[1] 0.1927(1)\footnotemark[2] & 0.1931(1) & 0.1930(1) & 0.1929(1) &  0.1932(1)  \\
\textrm{$y_{Ge}$ }   & 0.3857(1)\footnotemark[1] 0.3855(1)\footnotemark[2] & 0.3861(1) & 0.3864(1) & 0.3859(1) &  0.3865(1)  \\
\textrm{$\mu_{Cr}$ ($\mu_B$)}  & 0.64(6)\footnotemark[1] 0.90(7)\footnotemark[2] & 0.66(6) & 0.77(6) & 0.90(5)  &  1.40(5) \\
\textrm{$\chi^2$}   & 1.524\footnotemark[1] 1.559\footnotemark[2]  &  3.79     &  1.282    & 1.904  &  2.937    \\ 
\end{tabular}
\end{ruledtabular}
\footnotetext[1]{$\mu_{Cr}$ along the \textit{c} axis.}
\footnotetext[2]{$\mu_{Cr}$ parallel to the \textit{ab} plane.}
\end{table*}

NPD results for Ce$_{0.42}$La$_{0.58}$CrGe$_3$ and Ce$_{0.58}$La$_{0.43}$CrGe$_3$ measured at 5 K have revealed the same significant intensity enhancement for the (100) reflection when compared to the paramagnetic state at 100 K, as well as the small increase for (110) and (102) reflections (see Fig. \ref{fig5}). Consequently, we have analyzed these samples using the model with a single phase including structural and magnetic peaks with hexagonal \textit{P}6$_3$/\textit{mm'c'} symmetry. The Cr spins are aligned along the \textit{c} axis direction. Figures \ref{fig4} and \ref{fig5} displays the NPD and the magnetic peaks, and Table \ref{tab3} shows the structural and magnetic parameters obtained for all samples at 5 K.

In the case of Ce$_{0.81}$La$_{0.19}$CrGe$_3$ and CeCrGe$_3$ the NPD results at 5 K have shown not only a more evident (101) plane reflection than in the other samples when compared to that in the paramagnetic state measured at 100 K for Ce$_{0.81}$La$_{0.19}$CrGe$_3$ and 90 K for CeCrGe$_3$, but also an increase for the (100) and (102) reflections and possibly a very small contribution from the (002) and (200) reflections (see Fig. \ref{fig5}).
It is clear from Fig. \ref{fig5} that the intensity of the (100) reflection decreases whereas the intensity of the (101) reflection increases when the La concentration decreases. In this case, changes in the reflections could be related to a change in the magnetic moment value, change in the Cr spin orientation or, alternatively, a contribution from the Ce magnetic moment.

\begin{figure}
\includegraphics [scale=0.32]{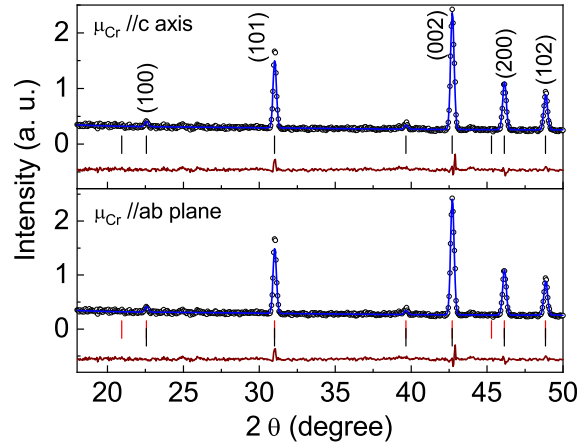}
\caption{NPD for CeCrGe$_3$ measured at 5 K with Rietveld refinement fit. Top, direction of Cr magnetic moment ($\mu_{Cr}$) along the \textit{c} axis. Bottom, direction of Cr magnetic moment parallel to the \textit{ab} plane. Blue lines are the Rietveld refinement fit and brown lines are the residual. In the top, vertical black lines indicate the theoretical magnetic peak positions with hexagonal \textit{P}6$_3$/\textit{mm'c'} symmetry. In the bottom, vertical black lines indicate the position of theoretical peaks for the structural phase and vertical red lines indicate the theoretical magnetic peak positions with 1-bar triclinic symmetry. All error bars represent 1 $\sigma$ and are shown, but they may be smaller than the symbol.}.
\label{fig6}
\end{figure}

Therefore, the refinement of NPD for CeCrGe$_3$ was analyzed with two models to describe the magnetic moments of the Cr ions.  The first model is the same used for the other samples, considering the alignment of Cr spins parallel to \textit{c} axis with $\mu_{Cr}$ = 0.636(63) $\mu_B$. The second model comprises two phases, one with only structural peaks and the other with only magnetic peaks in the 1-bar triclinic symmetry, resulting in the Cr spins aligned parallel to \textit{ab} plane with $\mu_{Cr}$ = 0.903(75) $\mu_B$. Fig. \ref{fig6} displays the Rietveld fit for each  model.
The magnetic moment of Ce atoms was not added to the model used to fit our NPD data because neither NPD results nor magnetization measurements, $\chi_{DC}$ and $\chi_{AC}$, revealed a coupling with the Ce ions. Thus, an ordered magnetic moment of Ce spins was not observed. We have tried to use the 1-bar triclinic symmetry to refine NPD data from Ce$_{0.81}$La$_{0.19}$CrGe$_3$ sample but the model that fit better was the same one used for the LaCrGe$_3$ sample with Cr spins parallel to the \textit{c} axis. 

Results from both models are presented in Table \ref{tab3}. The results of crystallographic parameters for CeCrGe$_3$ are in good agreement with those presented in previous studies\cite{Das2016} (\textit{a} = 6.1346(3) {\AA} and \textit{c} = 5.7083(4) {\AA} at 295 K). The refinements with Cr spins oriented either along the \textit{c} axis or parallel to the \textit{ab} plane lead to quite different values of the structural and magnetic parameters. It was not possible to observe a significant difference between the two models used in the fits. Using the values in Table \ref{tab3} for the Cr spins aligned along the \textit{c} axis for CeCrGe$_3$, it is clear that the values of the Cr magnetic moment, as well as the lattice parameters, gradually increase from \textit{x} = 0 to \textit{x} = 1, suggesting that this increase is caused by a chemical pressure as Ce ions are gradually substituted by La ions. The increase in the lattice parameters probably indicates that the Cr 3\textit{d} band is less hybridized with a consequent increase in the magnetic moment. On the other hand, when the CeCrGe$_3$ magnetic moment is parallel to the \textit{ab} plane, its value is higher than those for x = 0.19 and x = 0.43, suggesting an influence from Ce 4\textit{f} spin which was not accounted for.

\begin{figure}
\includegraphics [scale=0.3]{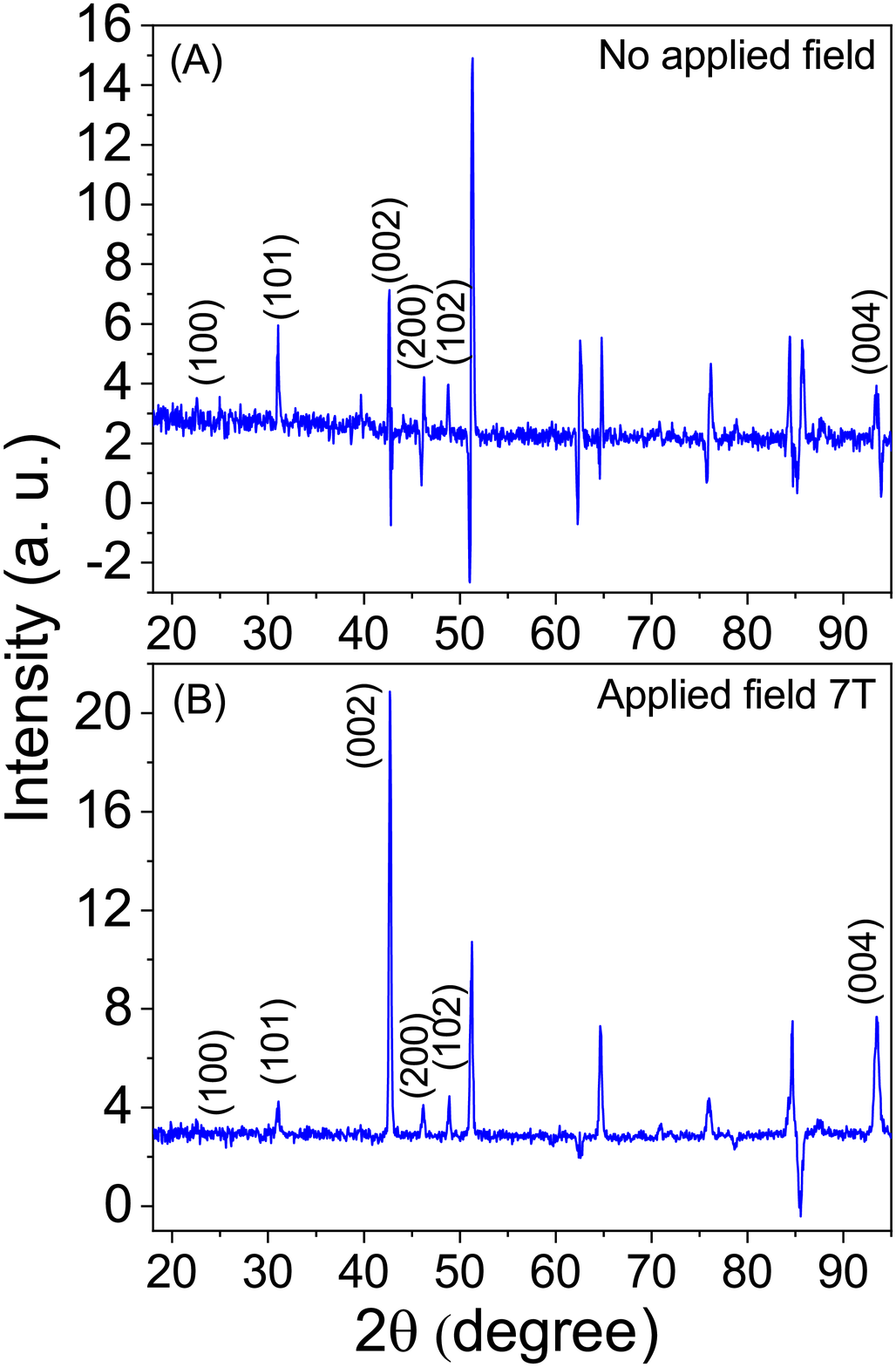}
\caption{CeCrGe$_3$ results: A) Magnetic intensity peaks from the difference between 5 K and 90 K and B) the difference between 0 T and 7 T at 5 K. All error bars represent 1 $\sigma$ and are shown, but they may be smaller than the symbol.}
\label{fig7}
\end{figure}

In order to show experimentally in which direction the Cr spins are aligned in this compound we performed a NPD study for CeCrGe$_3$ using a vertical field magnet system (7T VF) that applied an external magnetic field of 7 T. Fig. \ref{fig7} displays the pattern at 5 K with and without the application of the external magnetic field. From Fig. \ref{fig7} (B) one can observe the strong increase in the intensity of (002) and (004) plane reflections when compared to data from Fig. \ref{fig7} (A). This magnetic enhancement supports the idea that Cr spins in CeCrGe$_3$ are aligned parallel to the \textit{ab} plane. Even with the application of an external field, no evidence of a Ce spin contribution can be observed from the NPD measurements. According to Das \cite{Das2016} the magnetic contribution from the (101) reflection suggests ordering from Ce sublattice, but even with the application of a 7T external field this reflection did not increase (see Fig. \ref{fig7} (A) and Fig. \ref{fig7} (B)).

The changes in the intensity for the (100) and (101) reflections when La ions are gradually replaced by Ce ions (see Fig. \ref{fig5}) is additional evidence that the Cr spins are aligned in different directions in these compounds, as previously commented.

\subsubsection{Neutron scattering at BT-7 triple-axis spectrometer}

\begin{table}
\caption{\label{tab4}%
Magnetic Bragg peaks positions (2$\theta$) for the (100) reflection and transition temperature  (T$_C^{(Bt-7)}$) obtained from neutron scattering at BT-7 triple-axis spectrometer for Ce$_{(1-x)}$La$_{x}$CrGe$_3$ (\textit{x} = 0, 0.19, 0.43, 0.58, 1) series.}
\begin{ruledtabular}
\begin{tabular}{lcc}
         \textrm{Sample}&
         \textrm{2$\theta$ (degrees)}&
         \textrm{T$_C^{(Bt-7)}$ (K)}
        
\\       
\hline
\textrm{CeCrGe$_3$}                     & 25.5      & 77(1) \\
\textrm{Ce$_{0.81}$La$_{0.19}$CrGe$_3$} & 25.7      & 82(1) \\
\textrm{Ce$_{0.58}$La$_{0.43}$CrGe$_3$} & 25.7      & 84(1) \\
\textrm{Ce$_{0.42}$La$_{0.58}$CrGe$_3$ }& 25.8      & 92(1)\\
\textrm{LaCrGe$_3$}                     & 26        & 96(1) \\
\end{tabular}
\end{ruledtabular}
\end{table}

\begin{figure}[ht!]
\includegraphics [scale=0.4]{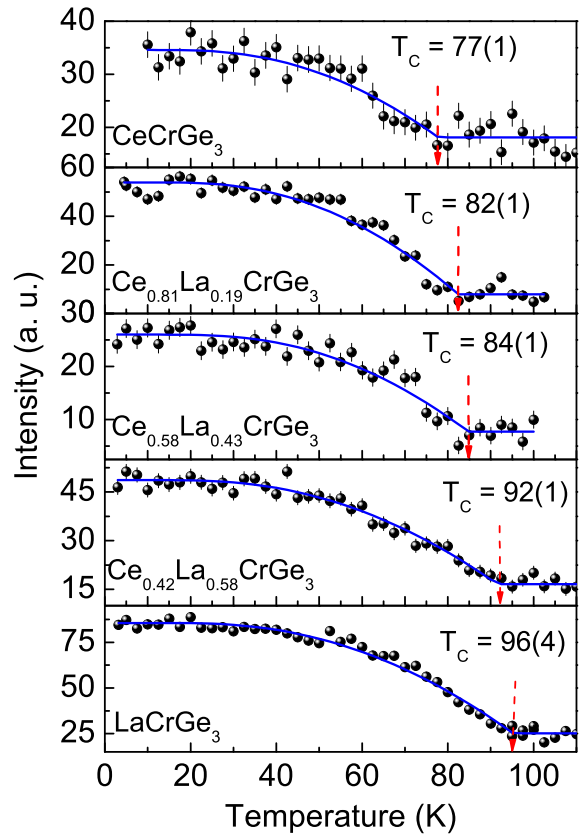}
\caption{Temperature dependence of the (100) integrated magnetic Bragg peak intensities. The blue solid curves show the mean field function fit. The vertical red dashed arrows indicate the $T_C$ obtained from the fit. All error bars are shown and represent 1 $\sigma$.  However, the error bars may be smaller than the symbol.}
\label{fig8}
\end{figure}

To gain a better understanding about the magnetic transition, the BT-7 triple-axis spectrometer data were collected using the magnetic Bragg peak located at the (100) peak for all samples in the series. The position for the (100) plane in each sample is shown in Table \ref{tab4}. Fig. \ref{fig8} shows the measurements of the integrated magnetic Bragg peaks intensities, which presents a smooth increase as the temperature decreases below T$_C^{(Bt-7)}$, indicating a second order magnetic transition. This behavior is associated with the Cr site magnetic order. At low temperatures the intensity exhibits the usual saturation that is typical of a conventional three-dimensional order parameter. Furthermore, the behavior for all samples show only one well-defined magnetic transition and it does not support the coexistence of both antiferromagnetic and ferromagnetic phases (as discussed in section A). Specifically, for LaCrGe$_3$ the integrated intensity does not show any evidence of a spin reorientation, as suggested by $\chi_{DC}$ around 80 K. Thus, the divergence between the ZFC and FC curves, and the bump in the ZFC curve for LaCrGe$_3$ (see Fig.\ref{fig1} (A)), can be better explained by magnetic domains, in agreement with $\chi_{AC}$ data.

The fits of mean field theory provide T$_C^{(Bt-7)}$ for all samples and are displayed at Table \ref{tab4}. The results from CeCrGe$_3$ and LaCrGe$_3$ show a difference when compared to those previously reported\cite{Das2014}:  $T_C$ = 73(1) K for CeCrGe$_3$ and  $T_C$ = 88(1) K for LaCrGe$_3$.

\section{\label{Disc}Discussion}
\begin{figure}[ht!]
\includegraphics [scale=0.22]{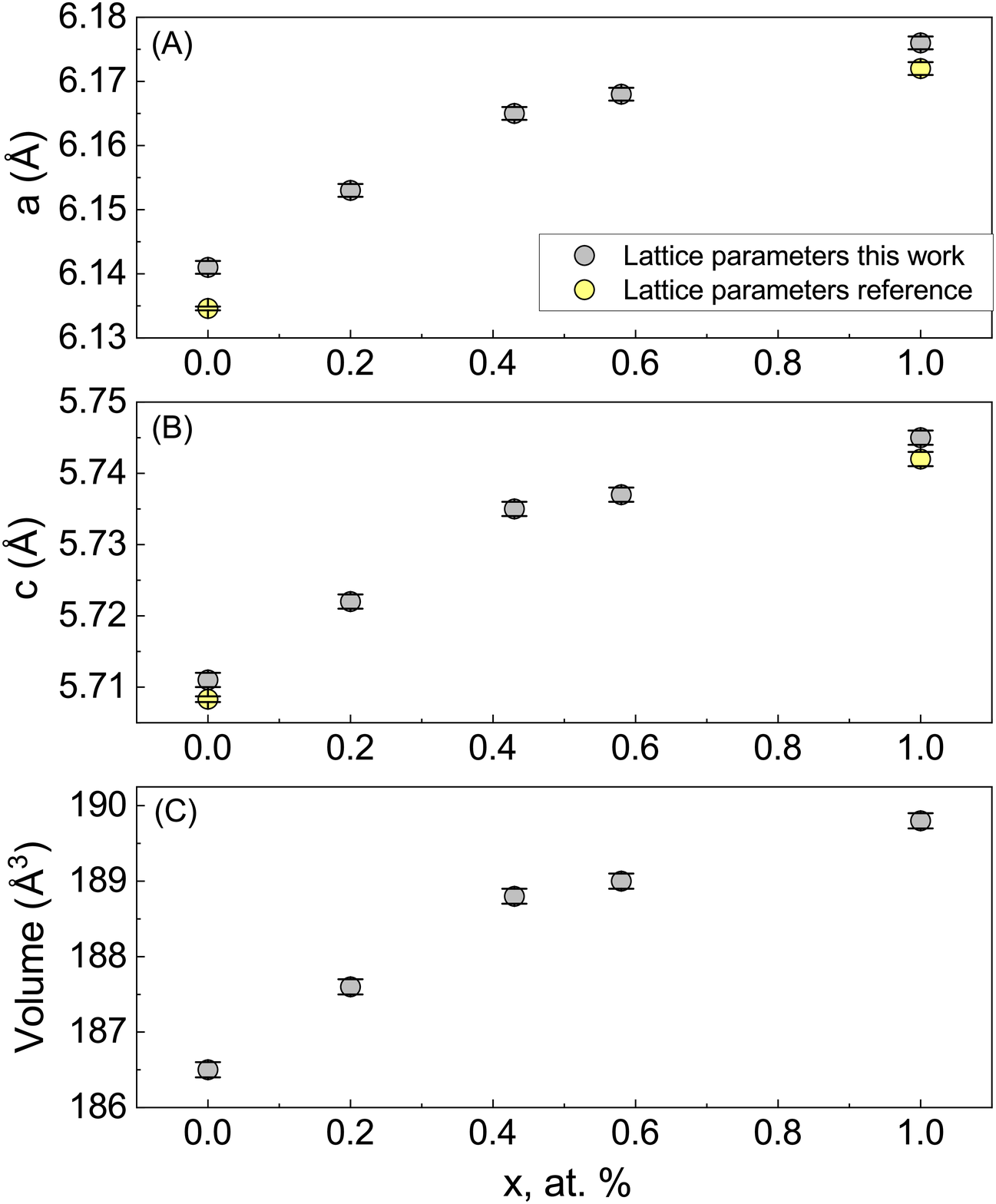}
\caption{Lattice parameters \textit{a}, \textit{c}, and unit cell volume at room temperature as a function of La-doping at room temperature. The error bars represent 1 $\sigma$.}.
\label{fig9}
\end{figure}

We have made a systematic investigation of the CeCrGe$_3$ structural and magnetic behavior, using magnetization measurements and neutron scattering techniques, as a function of doping La for Ce across the series to LaCrGe$_3$. The values of lattice parameters (\textit{a} and \textit{c}) and unit cell volume from NPD are displayed in Fig. \ref{fig9}. As expected, the gradual substitution of Ce by La leads to an increase in the lattice parameters and unit cell volume as a consequence of the well-known lanthanide expansion resulting in a bigger unit cell for LaCrGe$_3$, even though Ce ions have only one more 4\textit{f} electron. As shown in section \ref{Neutron}, we have obtained \textit{a}, \textit{c} and unit cell volume values for CeCrGe$_3$ and LaCrGe$_3$ at room temperature in good agreement with earlier published results.

\begin{figure}[ht!]
\includegraphics [scale=0.22]{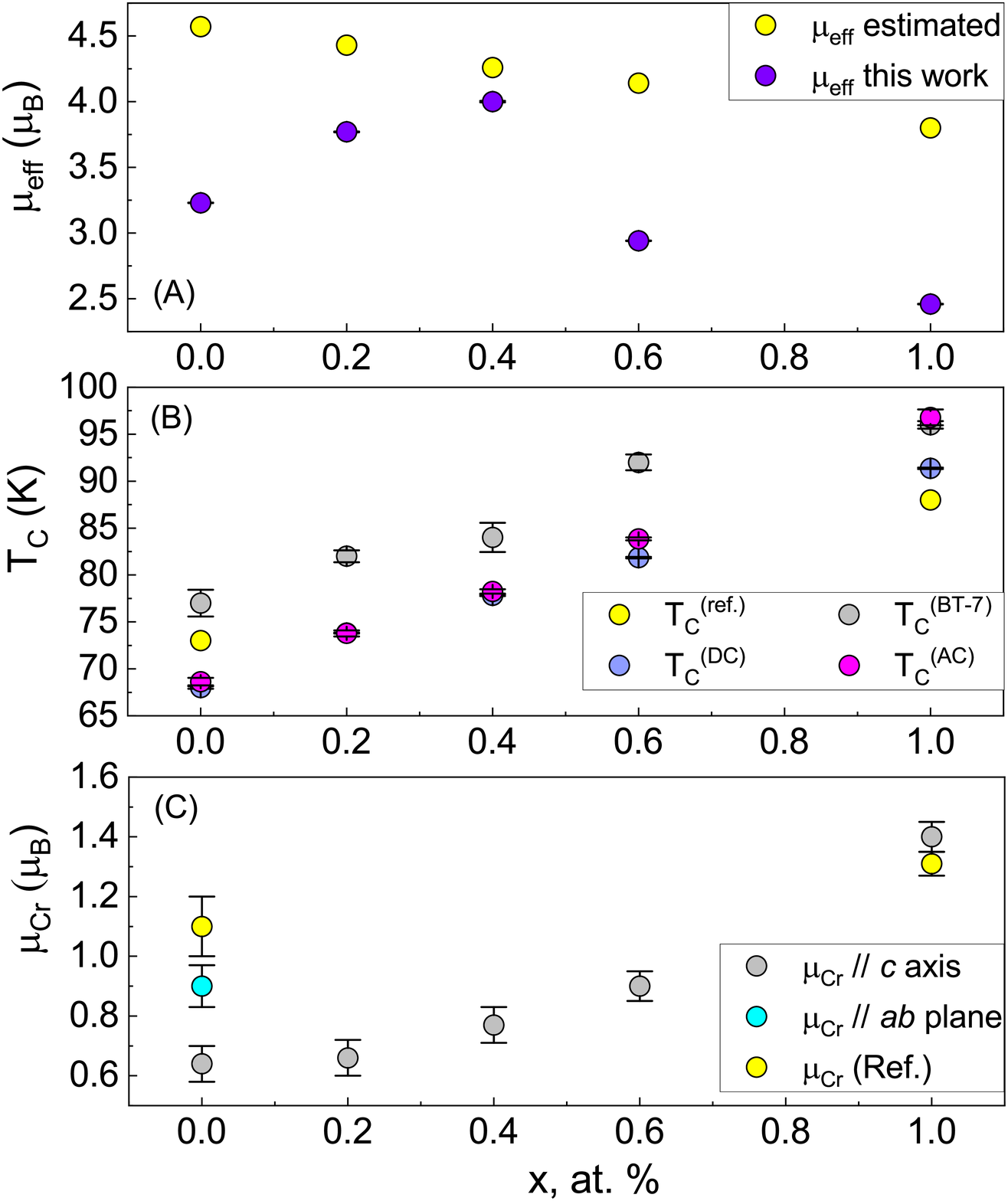}
\caption{(A) Effective magnetic moment ($\mu_{eff}$), (B) transition temperature ($T_C^{(DC)}$, $T_C^{(AC)}$ and $T_C^{(BT-7)}$) and (C) magnetic moment of the Cr atoms ($\mu_{Cr}$) from NPD as a function of La-doping (x) at room temperature. (The value $\mu_{Cr}$(Ref.) for CeCrGe$_3$ is parallel to the \textit{ab} plane and for LaCrGe$_3$ is parallel to the \textit{c} axis). Error bars where indicated represent one standard deviation.}
\label{fig10}
\end{figure}

 Fig.\ref{fig10} (A) shows the estimated $\mu_{eff}$ (y) and $\mu_{eff}$ obtained using the Curie-Weiss law for samples studied in this work as a function of La concentration. From the estimated $\mu_{eff}$ a linear behavior as a function of La concentration was expected. However, the results of  Curie–Weiss analysis in the Ce$_{(1-x)}$La$_x$CrGe$_3$ (x = 0, 0.19, 0.43 0.58) system show a non-linear behavior, suggesting a higher contribution from the Ce$^{3+}$ atoms for the x = 0.43 doped sample. Fig. \ref{fig10} (B) plots the variation of T$_C^{(DC)}$, T$_C^{(AC)}$ and T$_C^{(Bt-7)}$. The simplest explanation for this variation is due to the different methods of calculating T$_C$ not being identical.  Beyond this, there are two possible origins for this variation: spin-glass behavior or domain wall formation.  In spin glasses, where the relaxation time of the spins is much longer than the period of the AC frequency, this variation is to be expected between T$_C^{(DC)}$ and T$_C^{(AC)}$.  However, the AC susceptibility data do not show any variation with frequency and temperature.  In contrast, domain wall formation and then pining, especially just below T$_C$, could result in a similar variation, as the different methods may be more or less sensitive to the degree of domain formation.
 
 Additionally, comparing Fig. \ref{fig10} (A) and (C), the values of $\mu_{Cr}$ are much lower than $\mu_{eff}$. Such a difference suggests magnetic disorder reduces the spin polarization and consequently reduces the total ordered magnetic moment. In the case of  Ce$_{(1-x)}$La$_x$CrGe$_3$ (x = 0, 0.19, 0.43 0.58) this magnetic disorder could originate from the formation of domain walls at low temperature, where spin freezing may occur \cite{Nehla, Bocarsly}. Considering that a greater evidence of domain walls was seen in x = 0, x = 0.19 and x = 0.43 samples (see section \ref{Mag}), that domain structure originates from magneto-crystalline anisotropy, and is affected by defects in the crystal lattice   \cite{Levin} where the influence of Ce cannot be ruled out. For LaCrGe$_3$, as seen in the $\chi$' and $\chi$'' measurements, domain walls are also present and, similarly, at low temperatures, the spins freeze, causing this magnetic disorder  \cite{Nehla, Bocarsly}.
 
The prime motivation of this study was to determine the spin direction of Cr atoms in CeCrGe$_3$, because this information is not clear in the literature. To address this question, NPD measurements with an external magnetic field of 7 T have been carried out. The results show an increase in the intensity of the magnetic peak corresponding to the (002) reflection, which shows that the moments are aligning vertically, along the field direction in the ab plane. Additionally, in contrast to the statement by Das et.al.\citep{Das2016} on the coupling of Cr magnetic moments, the findings of the present study do not reveal the coupling to the Ce spins, even under an external magnetic field at the temperature of 5 K. But, from $\chi_{AC}$(T) results, there is a tendency of the Ce spins to order, suggesting an electron interaction of the Ce 4\textit{f} bands with Cr \textit{d} orbital are energetically promoting Cr spins to align in parallel with the \textit{ab} plane. We note that according to Nguyen, first-principles calculations show that LaCrGe$_3$ has a very strong peak just below the Fermi level, which is related to Cr \textit{d} orbital. Since the Cr spins delocalize with the application of an external pressure, the replacement of La by Ce provides a chemical pressure which causes a delocation of Cr spins, changing their alignment direction \cite{Nguyen2018}.

\section{Conclusions}

In summary, we have investigated the structural and magnetic behavior of Ce$_{(1-x)}$La$_x$CrGe$_3$ (x = 0, 0.19, 0.43, 0.58, 1). The quality of samples was checked by x-ray diffraction, magnetization measurements and neutron diffraction techniques. CeCrGe$_3$, Ce$_{0.58}$La$_{0.43}$CrGe$_3$ and LaCrGe$_3$ have no detectable impurity phases whereas Ce$_{0.81}$La$_{0.19}$CrGe$_3$ and Ce$_{0.42}$La$_{0.58}$CrGe$_3$ present a small second phase of non-magnetic La(Ce)Ge$_2$. Overall, results for CeCrGe$_3$ and LaCrGe$_3$ are consistent with other papers, and for x = 0.19, x = 0.43 and x = 0.58 results are in accordance with what is expected for the doped compounds of this family. Curie temperatures from different techniques, $T_C^{(DC)}$, $T_C^{(AC)}$ and $T_C^{(BT-7)}$, differ likely due to domain structure. The $\mu_{eff}$ agrees with the paramgnetic moment values as well as $\mu_{Cr}$ agrees with ferromagnetic moments values published previously  for CeCrGe$_3$ and LaCrGe$_3$ compounds. The unit cell volume enhancement as a function of La-doping concentration is directly related to the increase of $T_C$ and magnetic moment. LaCrGe$_3$ and the doped compounds with x = 0.19, x = 0.43 and x = 0.58 present Cr spins aligned along the \textit{c} axis direction, while CeCrGe$_3$ has the Cr spins aligned parallel to the \textit{ab} plane. There is no evidence of Ce spins ordering.

\begin{acknowledgments}
BBS thanks the financial support received from Coordena\c{c}\~{a}o de Aperfei\c{c}oamento de Pessoal de N\'{i}vel Superior (CAPES) by supporting postdoctoral research, grant number 88881.170115/2018-01. AWC kindly acknowledges the Conselho Nacional de Desenvolvimento Cient\'{i}fico e Tecnol\'{o}gico (CNPq) by support in a form of grant  304627/2017-8. Partial financial support for this work was provided by Funda\c{c}\~{a}o de Amparo a Pesquisa do Estado de S\~{a}o Paulo (FAPESP) under grant 2014/14001-1. We thank Craig Brown for helpful discussions about the refinements. The identification of any commercial product or trade name does not imply endorsement or recommendation by the National Institute of Standards and Technology.
\end{acknowledgments}

\bibliography{references_PRB.bib}

\providecommand{\noopsort}[1]{}\providecommand{\singleletter}[1]{#1}%
\begin{thebibliography}{24}%
\makeatletter
\providecommand \@ifxundefined [1]{%
 \@ifx{#1\undefined}
}%
\providecommand \@ifnum [1]{%
 \ifnum #1\expandafter \@firstoftwo
 \else \expandafter \@secondoftwo
 \fi
}%
\providecommand \@ifx [1]{%
 \ifx #1\expandafter \@firstoftwo
 \else \expandafter \@secondoftwo
 \fi
}%
\providecommand \natexlab [1]{#1}%
\providecommand \enquote  [1]{``#1''}%
\providecommand \bibnamefont  [1]{#1}%
\providecommand \bibfnamefont [1]{#1}%
\providecommand \citenamefont [1]{#1}%
\providecommand \href@noop [0]{\@secondoftwo}%
\providecommand \href [0]{\begingroup \@sanitize@url \@href}%
\providecommand \@href[1]{\@@startlink{#1}\@@href}%
\providecommand \@@href[1]{\endgroup#1\@@endlink}%
\providecommand \@sanitize@url [0]{\catcode `\\12\catcode `\$12\catcode
  `\&12\catcode `\#12\catcode `\^12\catcode `\_12\catcode `\%12\relax}%
\providecommand \@@startlink[1]{}%
\providecommand \@@endlink[0]{}%
\providecommand \url  [0]{\begingroup\@sanitize@url \@url }%
\providecommand \@url [1]{\endgroup\@href {#1}{\urlprefix }}%
\providecommand \urlprefix  [0]{URL }%
\providecommand \Eprint [0]{\href }%
\providecommand \doibase [0]{https://doi.org/}%
\providecommand \selectlanguage [0]{\@gobble}%
\providecommand \bibinfo  [0]{\@secondoftwo}%
\providecommand \bibfield  [0]{\@secondoftwo}%
\providecommand \translation [1]{[#1]}%
\providecommand \BibitemOpen [0]{}%
\providecommand \bibitemStop [0]{}%
\providecommand \bibitemNoStop [0]{.\EOS\space}%
\providecommand \EOS [0]{\spacefactor3000\relax}%
\providecommand \BibitemShut  [1]{\csname bibitem#1\endcsname}%
\let\auto@bib@innerbib\@empty
\bibitem [{\citenamefont {Manfrinetti}\ \emph {et~al.}(2005)\citenamefont
  {Manfrinetti}, \citenamefont {Dhar}, \citenamefont {Kulkarni},\ and\
  \citenamefont {Morozkin}}]{Manfrinetti2005}%
  \BibitemOpen
  \bibfield  {author} {\bibinfo {author} {\bibfnamefont {P.}~\bibnamefont
  {Manfrinetti}}, \bibinfo {author} {\bibfnamefont {S.~K.}\ \bibnamefont
  {Dhar}}, \bibinfo {author} {\bibfnamefont {R.}~\bibnamefont {Kulkarni}},\
  and\ \bibinfo {author} {\bibfnamefont {A.~V.}\ \bibnamefont {Morozkin}},\
  }\href@noop {} {\bibfield  {journal} {\bibinfo  {journal} {Solid State
  Comm.}\ }\textbf {\bibinfo {volume} {135}},\ \bibinfo {pages} {444} (\bibinfo
  {year} {2005})}\BibitemShut {NoStop}%
\bibitem [{\citenamefont {Bie}\ \emph {et~al.}(2007)\citenamefont {Bie},
  \citenamefont {Zelinska}, \citenamefont {Tkachuk},\ and\ \citenamefont
  {Mar}}]{Bie2007}%
  \BibitemOpen
  \bibfield  {author} {\bibinfo {author} {\bibfnamefont {H.}~\bibnamefont
  {Bie}}, \bibinfo {author} {\bibfnamefont {O.~Y.}\ \bibnamefont {Zelinska}},
  \bibinfo {author} {\bibfnamefont {A.~V.}\ \bibnamefont {Tkachuk}},\ and\
  \bibinfo {author} {\bibfnamefont {A.}~\bibnamefont {Mar}},\ }\href@noop {}
  {\bibfield  {journal} {\bibinfo  {journal} {Chem. Matter}\ }\textbf {\bibinfo
  {volume} {19}},\ \bibinfo {pages} {4613} (\bibinfo {year}
  {2007})}\BibitemShut {NoStop}%
\bibitem [{\citenamefont {Wang}\ \emph {et~al.}(2019)\citenamefont {Wang},
  \citenamefont {Guo}, \citenamefont {Bauer}, \citenamefont {Sidorov},
  \citenamefont {Zhao}, \citenamefont {Zhang}, \citenamefont {Zhou},
  \citenamefont {Wang}, \citenamefont {Cai}, \citenamefont {Yang},
  \citenamefont {Li}, \citenamefont {Sun}, \citenamefont {feng Yang},
  \citenamefont {Wu}, \citenamefont {Xiang}, \citenamefont {Thompson},\ and\
  \citenamefont {Sun}}]{Wang2019}%
  \BibitemOpen
  \bibfield  {author} {\bibinfo {author} {\bibfnamefont {H.}~\bibnamefont
  {Wang}}, \bibinfo {author} {\bibfnamefont {J.}~\bibnamefont {Guo}}, \bibinfo
  {author} {\bibfnamefont {E.~D.}\ \bibnamefont {Bauer}}, \bibinfo {author}
  {\bibfnamefont {V.~A.}\ \bibnamefont {Sidorov}}, \bibinfo {author}
  {\bibfnamefont {H.}~\bibnamefont {Zhao}}, \bibinfo {author} {\bibfnamefont
  {J.}~\bibnamefont {Zhang}}, \bibinfo {author} {\bibfnamefont
  {Y.}~\bibnamefont {Zhou}}, \bibinfo {author} {\bibfnamefont {Z.}~\bibnamefont
  {Wang}}, \bibinfo {author} {\bibfnamefont {S.}~\bibnamefont {Cai}}, \bibinfo
  {author} {\bibfnamefont {K.}~\bibnamefont {Yang}}, \bibinfo {author}
  {\bibfnamefont {A.}~\bibnamefont {Li}}, \bibinfo {author} {\bibfnamefont
  {P.}~\bibnamefont {Sun}}, \bibinfo {author} {\bibfnamefont {Y.}~\bibnamefont
  {feng Yang}}, \bibinfo {author} {\bibfnamefont {Q.}~\bibnamefont {Wu}},
  \bibinfo {author} {\bibfnamefont {T.}~\bibnamefont {Xiang}}, \bibinfo
  {author} {\bibfnamefont {J.~D.}\ \bibnamefont {Thompson}},\ and\ \bibinfo
  {author} {\bibfnamefont {L.}~\bibnamefont {Sun}},\ }\href@noop {} {\bibfield
  {journal} {\bibinfo  {journal} {Phys. Rev. B}\ }\textbf {\bibinfo {volume}
  {99}},\ \bibinfo {pages} {024504} (\bibinfo {year} {2019})}\BibitemShut
  {NoStop}%
\bibitem [{\citenamefont {Khan}\ \emph {et~al.}(2016)\citenamefont {Khan},
  \citenamefont {Yang}, \citenamefont {Mao}, \citenamefont {Du}, \citenamefont
  {Xu}, \citenamefont {Zhou}, \citenamefont {Zhang}, \citenamefont {Chen},\
  and\ \citenamefont {Fang}}]{Khan2016}%
  \BibitemOpen
  \bibfield  {author} {\bibinfo {author} {\bibfnamefont {R.}~\bibnamefont
  {Khan}}, \bibinfo {author} {\bibfnamefont {J.}~\bibnamefont {Yang}}, \bibinfo
  {author} {\bibfnamefont {H.~W.~Q.}\ \bibnamefont {Mao}}, \bibinfo {author}
  {\bibfnamefont {J.}~\bibnamefont {Du}}, \bibinfo {author} {\bibfnamefont
  {B.}~\bibnamefont {Xu}}, \bibinfo {author} {\bibfnamefont {Y.}~\bibnamefont
  {Zhou}}, \bibinfo {author} {\bibfnamefont {Y.}~\bibnamefont {Zhang}},
  \bibinfo {author} {\bibfnamefont {B.}~\bibnamefont {Chen}},\ and\ \bibinfo
  {author} {\bibfnamefont {M.}~\bibnamefont {Fang}},\ }\href@noop {} {\bibfield
   {journal} {\bibinfo  {journal} {Mater. Res. Express}\ }\textbf {\bibinfo
  {volume} {3}},\ \bibinfo {pages} {106101} (\bibinfo {year}
  {2016})}\BibitemShut {NoStop}%
\bibitem [{\citenamefont {Das}\ \emph {et~al.}(2015)\citenamefont {Das},
  \citenamefont {Bhattacharyya}, \citenamefont {Anand}, \citenamefont
  {Hillier}, \citenamefont {Taylor}, \citenamefont {Gruner}, \citenamefont
  {Geibel}, \citenamefont {Adroja},\ and\ \citenamefont {Hossain}}]{Das2015}%
  \BibitemOpen
  \bibfield  {author} {\bibinfo {author} {\bibfnamefont {D.}~\bibnamefont
  {Das}}, \bibinfo {author} {\bibfnamefont {A.}~\bibnamefont {Bhattacharyya}},
  \bibinfo {author} {\bibfnamefont {V.~K.}\ \bibnamefont {Anand}}, \bibinfo
  {author} {\bibfnamefont {A.~D.}\ \bibnamefont {Hillier}}, \bibinfo {author}
  {\bibfnamefont {J.~W.}\ \bibnamefont {Taylor}}, \bibinfo {author}
  {\bibfnamefont {T.}~\bibnamefont {Gruner}}, \bibinfo {author} {\bibfnamefont
  {C.}~\bibnamefont {Geibel}}, \bibinfo {author} {\bibfnamefont {D.~T.}\
  \bibnamefont {Adroja}},\ and\ \bibinfo {author} {\bibfnamefont
  {Z.}~\bibnamefont {Hossain}},\ }\href@noop {} {\bibfield  {journal} {\bibinfo
   {journal} {J. Phys.: Condensed Matter}\ }\textbf {\bibinfo {volume} {27}},\
  \bibinfo {pages} {016004} (\bibinfo {year} {2015})}\BibitemShut {NoStop}%
\bibitem [{\citenamefont {Lin}\ \emph {et~al.}(2013)\citenamefont {Lin},
  \citenamefont {Taufour}, \citenamefont {Bud'ko},\ and\ \citenamefont
  {Canfield}}]{Lin2013}%
  \BibitemOpen
  \bibfield  {author} {\bibinfo {author} {\bibfnamefont {X.}~\bibnamefont
  {Lin}}, \bibinfo {author} {\bibfnamefont {V.}~\bibnamefont {Taufour}},
  \bibinfo {author} {\bibfnamefont {S.~L.}\ \bibnamefont {Bud'ko}},\ and\
  \bibinfo {author} {\bibfnamefont {P.~C.}\ \bibnamefont {Canfield}},\
  }\href@noop {} {\bibfield  {journal} {\bibinfo  {journal} {Phys.\ Rev. B}\
  }\textbf {\bibinfo {volume} {88}},\ \bibinfo {pages} {094405} (\bibinfo
  {year} {2013})}\BibitemShut {NoStop}%
\bibitem [{\citenamefont {Taufour}\ \emph {et~al.}(2018)\citenamefont
  {Taufour}, \citenamefont {Kaluarachchi}, \citenamefont {Bud'ko},\ and\
  \citenamefont {Canfield}}]{Taufour2018}%
  \BibitemOpen
  \bibfield  {author} {\bibinfo {author} {\bibfnamefont {V.}~\bibnamefont
  {Taufour}}, \bibinfo {author} {\bibfnamefont {U.~S.}\ \bibnamefont
  {Kaluarachchi}}, \bibinfo {author} {\bibfnamefont {S.~L.}\ \bibnamefont
  {Bud'ko}},\ and\ \bibinfo {author} {\bibfnamefont {P.~C.}\ \bibnamefont
  {Canfield}},\ }\href@noop {} {\bibfield  {journal} {\bibinfo  {journal}
  {Physica B}\ }\textbf {\bibinfo {volume} {536}},\ \bibinfo {pages} {483}
  (\bibinfo {year} {2018})}\BibitemShut {NoStop}%
\bibitem [{\citenamefont {Taufour}\ \emph {et~al.}(2016)\citenamefont
  {Taufour}, \citenamefont {Kaluarachchi}, \citenamefont {Khasanov},\ and\
  \citenamefont {\textit{et. al}}}]{Taufour2016}%
  \BibitemOpen
  \bibfield  {author} {\bibinfo {author} {\bibfnamefont {V.}~\bibnamefont
  {Taufour}}, \bibinfo {author} {\bibfnamefont {U.~S.}\ \bibnamefont
  {Kaluarachchi}}, \bibinfo {author} {\bibfnamefont {R.}~\bibnamefont
  {Khasanov}},\ and\ \bibinfo {author} {\bibnamefont {\textit{et. al}}},\
  }\href@noop {} {\bibfield  {journal} {\bibinfo  {journal} {Phys. Rev. Lett.}\
  }\textbf {\bibinfo {volume} {117}},\ \bibinfo {pages} {037207} (\bibinfo
  {year} {2016})}\BibitemShut {NoStop}%
\bibitem [{\citenamefont {Das}\ \emph {et~al.}(2016)\citenamefont {Das},
  \citenamefont {Nandi}, \citenamefont {da~Silva}, \citenamefont {Adroja},\
  and\ \citenamefont {Hossain}}]{Das2016}%
  \BibitemOpen
  \bibfield  {author} {\bibinfo {author} {\bibfnamefont {D.}~\bibnamefont
  {Das}}, \bibinfo {author} {\bibfnamefont {S.}~\bibnamefont {Nandi}}, \bibinfo
  {author} {\bibfnamefont {I.}~\bibnamefont {da~Silva}}, \bibinfo {author}
  {\bibfnamefont {D.~T.}\ \bibnamefont {Adroja}},\ and\ \bibinfo {author}
  {\bibfnamefont {Z.}~\bibnamefont {Hossain}},\ }\href@noop {} {\bibfield
  {journal} {\bibinfo  {journal} {Phys.\ Rev. B}\ }\textbf {\bibinfo {volume}
  {94}},\ \bibinfo {pages} {174415} (\bibinfo {year} {2016})}\BibitemShut
  {NoStop}%
\bibitem [{\citenamefont {Das}\ \emph {et~al.}(2014)\citenamefont {Das},
  \citenamefont {Gruner}, \citenamefont {Pfau}, \citenamefont {Paramanik},
  \citenamefont {Burkhardt}, \citenamefont {Geibel},\ and\ \citenamefont
  {Hossain}}]{Das2014}%
  \BibitemOpen
  \bibfield  {author} {\bibinfo {author} {\bibfnamefont {D.}~\bibnamefont
  {Das}}, \bibinfo {author} {\bibfnamefont {T.}~\bibnamefont {Gruner}},
  \bibinfo {author} {\bibfnamefont {H.}~\bibnamefont {Pfau}}, \bibinfo {author}
  {\bibfnamefont {U.~B.}\ \bibnamefont {Paramanik}}, \bibinfo {author}
  {\bibfnamefont {U.}~\bibnamefont {Burkhardt}}, \bibinfo {author}
  {\bibfnamefont {C.}~\bibnamefont {Geibel}},\ and\ \bibinfo {author}
  {\bibfnamefont {Z.}~\bibnamefont {Hossain}},\ }\href@noop {} {\bibfield
  {journal} {\bibinfo  {journal} {J. Phys.: Condensed Matter}\ }\textbf
  {\bibinfo {volume} {26}},\ \bibinfo {pages} {106001} (\bibinfo {year}
  {2014})}\BibitemShut {NoStop}%
\bibitem [{\citenamefont {Nguyen}\ \emph {et~al.}(2018)\citenamefont {Nguyen},
  \citenamefont {Taufour}, \citenamefont {Bud'ko}, \citenamefont {Canfield},
  \citenamefont {Antropov}, \citenamefont {Wang},\ and\ \citenamefont
  {Ho}}]{Nguyen2018}%
  \BibitemOpen
  \bibfield  {author} {\bibinfo {author} {\bibfnamefont {M.~C.}\ \bibnamefont
  {Nguyen}}, \bibinfo {author} {\bibfnamefont {V.}~\bibnamefont {Taufour}},
  \bibinfo {author} {\bibfnamefont {S.~L.}\ \bibnamefont {Bud'ko}}, \bibinfo
  {author} {\bibfnamefont {P.~C.}\ \bibnamefont {Canfield}}, \bibinfo {author}
  {\bibfnamefont {V.~P.}\ \bibnamefont {Antropov}}, \bibinfo {author}
  {\bibfnamefont {C.-Z.}\ \bibnamefont {Wang}},\ and\ \bibinfo {author}
  {\bibfnamefont {K.-M.}\ \bibnamefont {Ho}},\ }\href@noop {} {\bibfield
  {journal} {\bibinfo  {journal} {Phys.\ Rev. B}\ }\textbf {\bibinfo {volume}
  {97}},\ \bibinfo {pages} {184401} (\bibinfo {year} {2018})}\BibitemShut
  {NoStop}%
\bibitem [{\citenamefont {Cadogan}\ \emph {et~al.}(2013)\citenamefont
  {Cadogan}, \citenamefont {Lemoine}, \citenamefont {Slater}, \citenamefont
  {Mar},\ and\ \citenamefont {Avdeev}}]{Cadogan2013}%
  \BibitemOpen
  \bibfield  {author} {\bibinfo {author} {\bibfnamefont {J.~M.}\ \bibnamefont
  {Cadogan}}, \bibinfo {author} {\bibfnamefont {P.}~\bibnamefont {Lemoine}},
  \bibinfo {author} {\bibfnamefont {B.~R.}\ \bibnamefont {Slater}}, \bibinfo
  {author} {\bibfnamefont {A.}~\bibnamefont {Mar}},\ and\ \bibinfo {author}
  {\bibfnamefont {M.}~\bibnamefont {Avdeev}},\ }\href@noop {} {\bibfield
  {journal} {\bibinfo  {journal} {Solid State Phenomena}\ }\textbf {\bibinfo
  {volume} {194}},\ \bibinfo {pages} {71} (\bibinfo {year} {2013})}\BibitemShut
  {NoStop}%
\bibitem [{\citenamefont {Lemoine}\ \emph {et~al.}(2013)\citenamefont
  {Lemoine}, \citenamefont {Cadogan}, \citenamefont {Slater}, \citenamefont
  {Mar},\ and\ \citenamefont {Avdeev}}]{Lemoine}%
  \BibitemOpen
  \bibfield  {author} {\bibinfo {author} {\bibfnamefont {P.}~\bibnamefont
  {Lemoine}}, \bibinfo {author} {\bibfnamefont {J.~M.}\ \bibnamefont
  {Cadogan}}, \bibinfo {author} {\bibfnamefont {B.~R.}\ \bibnamefont {Slater}},
  \bibinfo {author} {\bibfnamefont {A.}~\bibnamefont {Mar}},\ and\ \bibinfo
  {author} {\bibfnamefont {M.}~\bibnamefont {Avdeev}},\ }\href@noop {}
  {\bibfield  {journal} {\bibinfo  {journal} {J. Magn. Magn. Mater.}\ }\textbf
  {\bibinfo {volume} {325}},\ \bibinfo {pages} {135} (\bibinfo {year}
  {2013})}\BibitemShut {NoStop}%
\bibitem [{\citenamefont {Binder}\ and\ \citenamefont {Young}(1986)}]{Binder}%
  \BibitemOpen
  \bibfield  {author} {\bibinfo {author} {\bibfnamefont {K.}~\bibnamefont
  {Binder}}\ and\ \bibinfo {author} {\bibfnamefont {A.~P.}\ \bibnamefont
  {Young}},\ }\href@noop {} {\bibfield  {journal} {\bibinfo  {journal} {Rev.
  Mod. Phys.}\ }\textbf {\bibinfo {volume} {58}},\ \bibinfo {pages} {801}
  (\bibinfo {year} {1986})}\BibitemShut {NoStop}%
\bibitem [{\citenamefont {Bocarsly}\ \emph {et~al.}(2019)\citenamefont
  {Bocarsly}, \citenamefont {Heikes}, \citenamefont {Brown}, \citenamefont
  {Wilson},\ and\ \citenamefont {Seshadri}}]{Bocarsly}%
  \BibitemOpen
  \bibfield  {author} {\bibinfo {author} {\bibfnamefont {J.~D.}\ \bibnamefont
  {Bocarsly}}, \bibinfo {author} {\bibfnamefont {C.}~\bibnamefont {Heikes}},
  \bibinfo {author} {\bibfnamefont {C.~M.}\ \bibnamefont {Brown}}, \bibinfo
  {author} {\bibfnamefont {S.~D.}\ \bibnamefont {Wilson}},\ and\ \bibinfo
  {author} {\bibfnamefont {R.}~\bibnamefont {Seshadri}},\ }\href@noop {}
  {\bibfield  {journal} {\bibinfo  {journal} {Phys. Rev. Mat.}\ }\textbf
  {\bibinfo {volume} {3}},\ \bibinfo {pages} {014402} (\bibinfo {year}
  {2019})}\BibitemShut {NoStop}%
\bibitem [{\citenamefont {Nehla}\ \emph {et~al.}(2019)\citenamefont {Nehla},
  \citenamefont {Kareri}, \citenamefont {Gupt}, \citenamefont {Hester},
  \citenamefont {Babu}, \citenamefont {Ulrich},\ and\ \citenamefont
  {Dhakai}}]{Nehla}%
  \BibitemOpen
  \bibfield  {author} {\bibinfo {author} {\bibfnamefont {P.}~\bibnamefont
  {Nehla}}, \bibinfo {author} {\bibfnamefont {Y.}~\bibnamefont {Kareri}},
  \bibinfo {author} {\bibfnamefont {G.~D.}\ \bibnamefont {Gupt}}, \bibinfo
  {author} {\bibfnamefont {J.}~\bibnamefont {Hester}}, \bibinfo {author}
  {\bibfnamefont {P.~D.}\ \bibnamefont {Babu}}, \bibinfo {author}
  {\bibfnamefont {C.}~\bibnamefont {Ulrich}},\ and\ \bibinfo {author}
  {\bibfnamefont {R.~S.}\ \bibnamefont {Dhakai}},\ }\href@noop {} {\bibfield
  {journal} {\bibinfo  {journal} {Phys.\ Rev. B}\ }\textbf {\bibinfo {volume}
  {100}},\ \bibinfo {pages} {144444} (\bibinfo {year} {2019})}\BibitemShut
  {NoStop}%
\bibitem [{\citenamefont {Eremenko}\ \emph {et~al.}(1971)\citenamefont
  {Eremenko}, \citenamefont {Shi}, \citenamefont {Buyanov},\ and\ \citenamefont
  {Batalin}}]{Eremenko1971}%
  \BibitemOpen
  \bibfield  {author} {\bibinfo {author} {\bibfnamefont {V.~N.}\ \bibnamefont
  {Eremenko}}, \bibinfo {author} {\bibfnamefont {Z.~K.}\ \bibnamefont {Shi}},
  \bibinfo {author} {\bibfnamefont {Y.~I.}\ \bibnamefont {Buyanov}},\ and\
  \bibinfo {author} {\bibfnamefont {V.~G.}\ \bibnamefont {Batalin}},\
  }\href@noop {} {\bibfield  {journal} {\bibinfo  {journal} {Poroshkovaya
  Metallurgiya}\ }\textbf {\bibinfo {volume} {8}},\ \bibinfo {pages} {82}
  (\bibinfo {year} {1971})}\BibitemShut {NoStop}%
\bibitem [{\citenamefont {Santoro}(2001)}]{Santoro2001}%
  \BibitemOpen
  \bibfield  {author} {\bibinfo {author} {\bibfnamefont {A.}~\bibnamefont
  {Santoro}},\ }\href@noop {} {\bibfield  {journal} {\bibinfo  {journal} {J.
  Res. Natl. Inst. Stand. Technol.}\ }\textbf {\bibinfo {volume} {106}},\
  \bibinfo {pages} {921} (\bibinfo {year} {2001})}\BibitemShut {NoStop}%
\bibitem [{\citenamefont {Larson}\ and\ \citenamefont {Dreele}(2000)}]{GSAS}%
  \BibitemOpen
  \bibfield  {author} {\bibinfo {author} {\bibfnamefont {A.~C.}\ \bibnamefont
  {Larson}}\ and\ \bibinfo {author} {\bibfnamefont {R.~V.}\ \bibnamefont
  {Dreele}},\ }\href@noop {} {\emph {\bibinfo {title} {General Structure
  Analysis System (GSAS)}}},\ \bibinfo {organization} {Los Alamos National
  Laboratory Report LAUR},\ \bibinfo {address} {New Mexico, USA} (\bibinfo
  {year} {2000})\BibitemShut {NoStop}%
\bibitem [{\citenamefont {Lynn}\ \emph {et~al.}(2012)\citenamefont {Lynn},
  \citenamefont {Chen}, \citenamefont {Chang}, \citenamefont {Zhao},
  \citenamefont {Chi}, \citenamefont {W.~Ratcliff}, \citenamefont {Ueland},\
  and\ \citenamefont {Erwin}}]{Lynn2012}%
  \BibitemOpen
  \bibfield  {author} {\bibinfo {author} {\bibfnamefont {J.~W.}\ \bibnamefont
  {Lynn}}, \bibinfo {author} {\bibfnamefont {Y.}~\bibnamefont {Chen}}, \bibinfo
  {author} {\bibfnamefont {S.}~\bibnamefont {Chang}}, \bibinfo {author}
  {\bibfnamefont {Y.}~\bibnamefont {Zhao}}, \bibinfo {author} {\bibfnamefont
  {S.}~\bibnamefont {Chi}}, \bibinfo {author} {\bibfnamefont {I.}~\bibnamefont
  {W.~Ratcliff}}, \bibinfo {author} {\bibfnamefont {B.~G.}\ \bibnamefont
  {Ueland}},\ and\ \bibinfo {author} {\bibfnamefont {R.~W.}\ \bibnamefont
  {Erwin}},\ }\href@noop {} {\bibfield  {journal} {\bibinfo  {journal} {J. Res.
  Natl Inst. Stand. Technol.}\ }\textbf {\bibinfo {volume} {117}},\ \bibinfo
  {pages} {61} (\bibinfo {year} {2012})}\BibitemShut {NoStop}%
\bibitem [{\citenamefont {Azuah}\ \emph {et~al.}(2009)\citenamefont {Azuah},
  \citenamefont {Kneller}, \citenamefont {Qiu}, \citenamefont
  {Tregenna-Piggott}, \citenamefont {Brown}, \citenamefont {Copley},\ and\
  \citenamefont {Dimeo}}]{DAVE}%
  \BibitemOpen
  \bibfield  {author} {\bibinfo {author} {\bibfnamefont {R.}~\bibnamefont
  {Azuah}}, \bibinfo {author} {\bibfnamefont {L.}~\bibnamefont {Kneller}},
  \bibinfo {author} {\bibfnamefont {Y.}~\bibnamefont {Qiu}}, \bibinfo {author}
  {\bibfnamefont {P.}~\bibnamefont {Tregenna-Piggott}}, \bibinfo {author}
  {\bibfnamefont {C.}~\bibnamefont {Brown}}, \bibinfo {author} {\bibfnamefont
  {J.}~\bibnamefont {Copley}},\ and\ \bibinfo {author} {\bibfnamefont
  {R.}~\bibnamefont {Dimeo}},\ }\href@noop {} {\bibfield  {journal} {\bibinfo
  {journal} {J. Res. Natl Inst. Stand. Technol.}\ }\textbf {\bibinfo {volume}
  {114}},\ \bibinfo {pages} {341} (\bibinfo {year} {2009})}\BibitemShut
  {NoStop}%
\bibitem [{\citenamefont {Amarotti}\ and\ \citenamefont
  {Fournier}(1984)}]{Amarotti1984}%
  \BibitemOpen
  \bibfield  {author} {\bibinfo {author} {\bibfnamefont {G.}~\bibnamefont
  {Amarotti}}\ and\ \bibinfo {author} {\bibfnamefont {J.~M.}\ \bibnamefont
  {Fournier}},\ }\href@noop {} {\bibfield  {journal} {\bibinfo  {journal} {J.
  Magn. Magn. Mater.}\ }\textbf {\bibinfo {volume} {43}},\ \bibinfo {pages}
  {L217} (\bibinfo {year} {1984})}\BibitemShut {NoStop}%
\bibitem [{\citenamefont {Levin}\ \emph {et~al.}(2001)\citenamefont {Levin},
  \citenamefont {Pecharsky},\ and\ \citenamefont {Gschneidner}}]{Levin}%
  \BibitemOpen
  \bibfield  {author} {\bibinfo {author} {\bibfnamefont {E.~M.}\ \bibnamefont
  {Levin}}, \bibinfo {author} {\bibfnamefont {V.~K.}\ \bibnamefont
  {Pecharsky}},\ and\ \bibinfo {author} {\bibfnamefont {K.~A.}\ \bibnamefont
  {Gschneidner}},\ }\href@noop {} {\bibfield  {journal} {\bibinfo  {journal}
  {J. Appl. Phys.}\ }\textbf {\bibinfo {volume} {90}},\ \bibinfo {pages} {6255}
  (\bibinfo {year} {2001})}\BibitemShut {NoStop}%
\bibitem [{\citenamefont {Aslibeiki}\ \emph {et~al.}(2009)\citenamefont
  {Aslibeiki}, \citenamefont {Kameli},\ and\ \citenamefont
  {Salamati}}]{Aslibeiki}%
  \BibitemOpen
  \bibfield  {author} {\bibinfo {author} {\bibfnamefont {B.}~\bibnamefont
  {Aslibeiki}}, \bibinfo {author} {\bibfnamefont {P.}~\bibnamefont {Kameli}},\
  and\ \bibinfo {author} {\bibfnamefont {H.}~\bibnamefont {Salamati}},\
  }\href@noop {} {\bibfield  {journal} {\bibinfo  {journal} {Solid State
  Communications.}\ }\textbf {\bibinfo {volume} {149}},\ \bibinfo {pages}
  {1274} (\bibinfo {year} {2009})}\BibitemShut {NoStop}%
\end{thebibliography}%

\end{document}